\documentclass[article]{aa}  
\usepackage{graphicx}
\usepackage{txfonts}
\usepackage{color}
\usepackage{pdflscape}
\definecolor{aipred}{rgb}{0.63,0.0,0.16}
\definecolor{aipdarkblue}{rgb}{0.0,0.25,0.47}
\usepackage[colorlinks=true,urlcolor=blue,citecolor=aipdarkblue]{hyperref}
\newcommand{\ha}{H$\alpha$}
\newcommand{\hb}{H$\beta$}
\newcommand{\hg}{H$\gamma$}

\newcommand{\oiii}{[\ion{O}{iii}]$\lambda5007$}
\newcommand{\heii}{\ion{He}{ii} $\lambda4686$}
\newcommand{\siidoublet}{[\ion{S}{ii}]$\lambda6716,6731$}

\newcommand{\nii}{[\ion{N}{ii}]$\lambda6584$}

\newcommand{\sii}{[\ion{S}{ii}]$\lambda6716$}
\newcommand{\siiright}{[\ion{S}{ii}]$\lambda6731$}
\newcommand{\siisum}{[\ion{S}{ii}]$\lambda\lambda6716,6731$}
\newcommand{\siirat}{[\ion{S}{ii}]$\lambda6731/6716$}

\newcommand{\hii}{\ion{H}{ii}}
\usepackage[]{hyperref}
\begin{document}

   \title{MUSE crowded field 3D spectroscopy in NGC\,300}

   \subtitle{IV. Planetary nebula luminosity function}

   \author{Azlizan A. Soemitro \inst{1,2}
          \and
          Martin M. Roth \inst{1,2}
          \and
          Peter M. Weilbacher \inst{1}
          \and
          Robin Ciardullo  \inst{3,4}
          \and
          George H. Jacoby \inst{5}
          \and
          Ana Monreal-Ibero \inst{6}
          \and
          Norberto Castro\inst{1} 
          \and
          Genoveva Micheva\inst{1}
          }

   \institute{Leibniz-Institut für Astrophysik Potsdam (AIP),
              An der Sternwarte 16, 14482 Potsdam, Germany\\
              \email{asoemitro@aip.de}
         \and
            Institut für Physik und Astronomie, Universität Potsdam, Karl-Liebknecht-Str. 24/25, 14476 Potsdam, Germany
         \and
            Department of Astronomy \& Astrophysics, The Pennsylvania State University, University Park, PA 16802, USA
         \and
            Institute for Gravitation and the Cosmos, The Pennsylvania State University, University Park, PA 16802, USA
         \and
            NSF’s NOIRLab, 950 N. Cherry Ave., Tucson, AZ 85719, USA
         \and
            Leiden Observatory, Leiden University, PO Box 9513, 2300 RA Leiden, The Netherlands
             }

   \date{Received ; accepted }

  \abstract
   {}
   {We perform a deep survey of planetary nebulae (PNe) in the spiral galaxy NGC\,300 to construct its planetary nebula luminosity function (PNLF). We aim to derive the distance using the PNLF and to probe the characteristics of the most luminous PNe.}
   {We analyse 44 fields observed with MUSE at the VLT, covering a total area of $\sim11$ kpc$^2$. We find \oiii~sources using the differential emission line filter (DELF) technique. We identify PNe through spectral classification using the aid of the BPT-diagram. The PNLF distance is derived using the maximum likelihood estimation technique. For the more luminous PNe, we also measure their extinction using the Balmer decrement. We estimate the luminosity and effective temperature of the central stars of the luminous PNe, based on estimates of the excitation class and the assumption of optically thick nebulae.}
   {We identify 107 PNe and derive a most-likely distance modulus $(m-M)_0 = 26.48^{+0.11}_{-0.26}$ ($d = 1.98^{+0.10}_{-0.23}$ Mpc). We find that the PNe at the PNLF cut-off exhibit relatively low extinction, with some high extinction cases caused by local dust lanes. We present the lower limit luminosities and effective temperatures of the central stars for some of the brighter PNe. 
   We also identify a few Type I PNe that come from a young population with progenitor masses $>2.5 \, M_\odot$, however do not populate the PNLF cut-off.}
   {The spatial resolution and spectral information of MUSE allow precise PN classification and photometry. These capabilities also enable us to resolve possible contamination by diffuse gas and dust, improving the accuracy of the PNLF distance to NGC\,300.}

   \keywords{galaxies: stellar content --
                planetary nebulae: general -- galaxies: luminosity function, mass function -- distance scale -- stars: AGB and post-AGB
               }

\maketitle

%

\section{Introduction}

The planetary nebula luminosity function (PNLF) is a distance determination method with a precision and accuracy that is comparable to those of the tip of the red giant branch (TRGB) and Cepheid methods  \citep{1989ApJ...339...39J, 1989ApJ...339...53C, 2010PASA...27..149C, 2012Ap&SS.341..151C, 2021arXiv210501982R}. Using evidence from narrow-band photometric surveys in \oiii, \citet{1989ApJ...339...53C} have shown that the magnitude distribution of planetary nebulae (PNe) for a given galaxy follows an empirical power law defined as 
\begin{equation}
    N(M) \propto e^{0.307M}\{1-e^{3(M^*-M)}\}
    \label{eq:pnlf}
\end{equation}
where the brightest PN at the cut-off has an absolute magnitude of $M^*=-4.53 \pm 0.06$ with a possible minor dependency on metallicity \citep{1989ApJ...339...39J, 1992ApJ...389...27D,2002ApJ...577...31C,2012Ap&SS.341..151C}. While a number of different formulations have been developed to model the various shapes of the PNLF at fainter magnitudes \citep{2015A&A...575A...1R, 2013A&A...558A..42L, 2019A&A...624A.132B,2021A&A...647A.130B}, such faint-end variation do not affect the definition of the PNLF's bright end cut-off, which is the critical feature for distance determinations \citep{2021A&A...653A.167S, 2022FrASS...9.6326C}. 

Until the early 2010s, most PNLF distance measurements were obtained using 4-meter class telescopes and narrow-band interference filters, and as a result,  the method has been traditionally limited to distances of $\sim 20$ Mpc \citep{1990ApJ...356..332J, 2010PASA...27..149C, 2012Ap&SS.341..151C, 2022FrASS...9.6326C}. Although 8-meter class telescopes were available and even observed PNe at the Coma cluster \citep[$\sim$ 100 Mpc,][]{2005ApJ...621L..93G}, most of the instruments had wider bandpass filters, which increased the inclusion of sky background signal.
This limited the PN detection sensitivity, that was necessary to significantly improve the distance range of the PNLF \citep{2022FrASS...9.6326C}. This situation has now changed due to the use of the Multi Unit Spectroscopic Explorer \citep[MUSE,][]{2010SPIE.7735E..08B} integral-field spectrograph on the 8.2-meter Very Large Telescope to survey PNe in distant systems \citep{2020A&A...637A..62S,2021A&A...653A.167S,2021arXiv210501982R,2022MNRAS.511.6087S}.  In fact, \citet{2021arXiv210501982R} have shown that by using a differential emission line filter technique on MUSE data, PNLF measurements are now possible out to distances of $\sim 40$ Mpc under excellent seeing condition and with the aid of the adaptive optics system. This is mainly due to the narrow effective bandpass of MUSE, that is five times narrower than the typical narrow-band filters, which can substantially suppressed the background sky noise \citep{2021arXiv210501982R}.

Previous PNLF studies of late type galaxies using \oiii~narrow-band filters were also hampered by the possible confusion with supernova remnants (SNRs) or \hii~regions \citep{2008ApJ...683..630H, 2009ApJ...703..894H, 2010PASA...27..129F}. While \ha~narrow-band image allows the exclusion of \hii~regions, it cannot be used to exclude the SNRs, whose classification typically rely on the \siidoublet~lines. In M31 and M33, \citet{Davis:2018aa} found that the SNR contamination does not change the shape nor the position of the PNLF cutoff. In contrary, \citet{2017ApJ...834..174K} have shown with MUSE observations of NGC\,628 that the presence of SNR contaminants can affect the bright cut-off. However, \citet{2022MNRAS.511.6087S} demonstrated that the latter study had an issue with the background subtraction in \ha, which affected the classification. Their reanalysis concluded that the PNLF cutoff was indeed unaffected by the contaminants. Nevertheless, the possible contamination by SNRs is relevant for the investigation of the faint end of the PNLF. The impressive spatial resolution and spectroscopic capability of the MUSE instrument allows the instant identification of interlopers, even in star forming disk galaxies \citep{2017ApJ...834..174K,  2018A&A...618A...3R, 2021arXiv210501982R, 2022MNRAS.511.6087S}.  


NGC\,300 is a spiral galaxy in the foreground of the Sculptor group. Being fairly isolated from its neighbouring galaxies \citep{2003A&A...404...93K} and close in distance \citep{2005ApJ...628..695G, 2006ApJ...638..766R, 2007ApJ...661..815R} makes it interesting for studying star formation histories and galactic evolution \citep{2007ApJ...658.1006M, 2008ApJ...681..269K, 2009ApJ...699.1541B, 2010ApJ...712..858G, 2020A&A...640L..19J}. A previous PNLF study, \citet{1996A&A...306....9S} identified 34 PNe, a small number that was not ideal to make a proper PNLF and therefore opted the cumulative PNLF to derive the distance by using the LMC as a yardstick. More recently, \citet{2012A&A...547A..78P} observed 104 PN candidates using narrow-band imaging from the central and the eastern outskirt region to construct the PNLF, with a follow-up spectroscopy for the brighter candidates \citep{2013A&A...552A..12S}. In Paper I \citep{2018A&A...618A...3R}, seven $1\arcmin \times 1\arcmin$ MUSE fields in the central region of NGC\,300 were observed with the goal of resolving stellar populations in crowded fields of nearby galaxies, from which they discovered 45 PN candidates. Again, this number was too small to create a useful PNLF, since the sample spans a very wide magnitude range of $22 \lesssim m_{5007} \lesssim 29$. In the present work, using publicly available archival data from \citet{2020ApJ...891...25M, 2021MNRAS.508.5425M} -- or ML20 -- and 2 additional MUSE-GTO fields, we expand the observed area from 7 to 44 MUSE fields in order to detect more PNe and obtain a PNLF distance to NGC\,300 using integral field spectroscopy.    


Our lack of a complete understanding of the underlying physics behind the invariance of the PNLF cut-off has prevented the PNLF technique to become a primary standard candle \citep{2010PASA...27..149C, 2012Ap&SS.341..151C}. Although simulations have provided an impression of the physical properties of the most luminous PNe \citep{1989ApJ...339...39J, 1990ApJ...357..140D, 1991ApJ...367..115D, 1997A&A...321..898M, 2008ApJ...681..325M, 2007A&A...473..467S, 2010A&A...523A..86S, 2019AAS...23315001V}, an observational characterisation is still limited to the LMC \citep{1991ApJ...367..115D, 1992ApJ...389...27D, 2010MNRAS.405.1349R, 2010PASA...27..187R}, and M31 \citep{2012ApJ...753...12K, 2018ApJ...863..189D, 2022A&A...657A..71G}. If the most luminous PNe at the PNLF cut-off have indeed originated from a single-star stellar evolution, then placing the central stars in the HR-diagram will provide insights into the underlying stellar population, and also the nature of the cut-off itself. Using the data quality that MUSE offers, we aim to constrain the luminosity and effective temperature of the central stars for some of the bright PNe to understand their origin and expand our understanding of PNe beyond the Local Group. 

The structure of this paper is as follows: details on observations and data reduction are described in Section 2. The data analysis regarding the PN detection and classification, the literature comparison of the PN number, the \oiii~photometry, and the measurement of the Balmer decrement is explained in Section 3. The resulting luminosity function and the distance measurement are presented in Section 4. The discussion and the implications of this work follow in Section 5. Lastly, the conclusions are given in Section 6.

  \begin{figure*}
   \centering
   \includegraphics[width=\hsize]{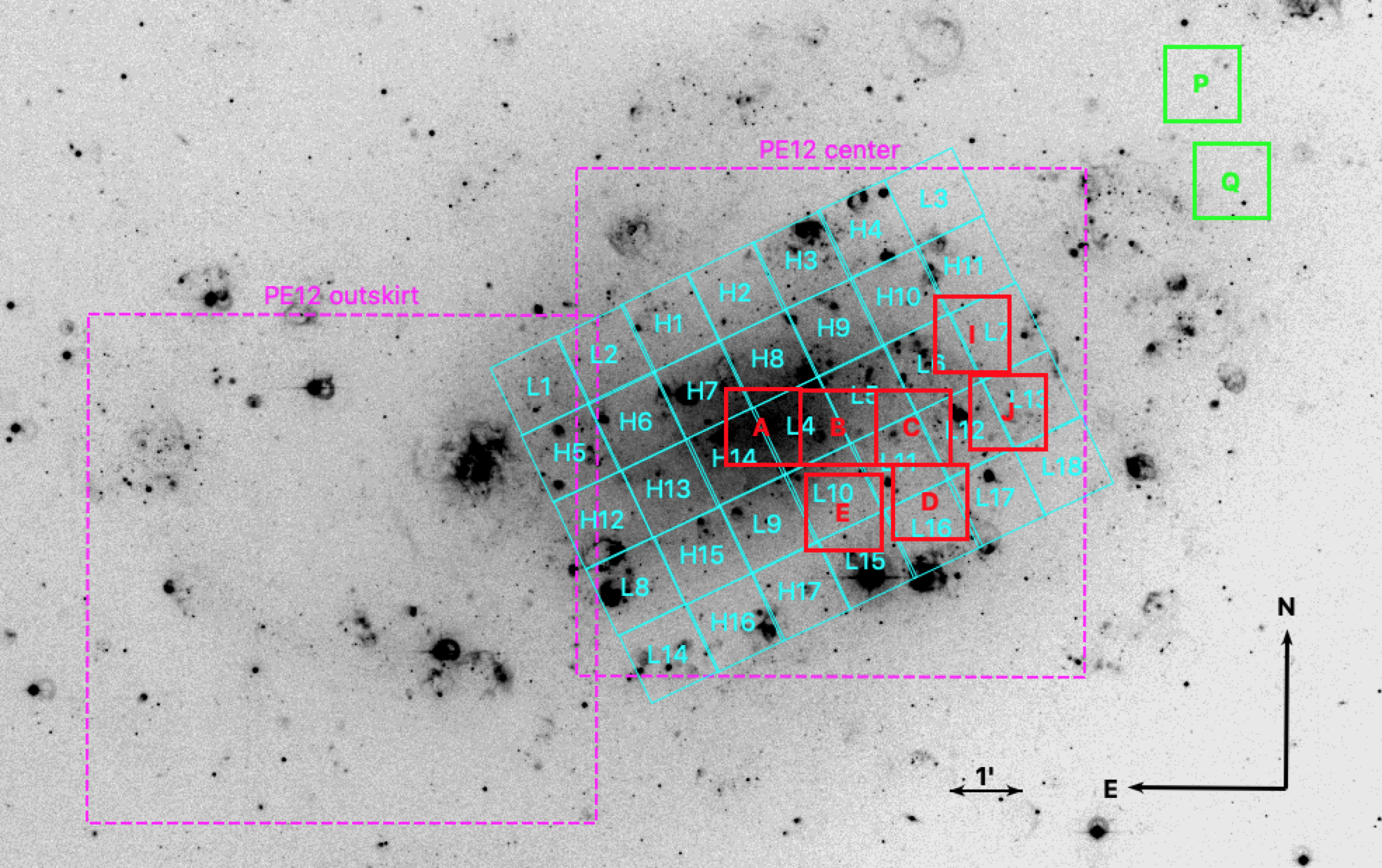}
      \caption{MUSE fields of NGC\,300 are marked with red, green, and cyan. The red fields are MUSE-GTO data from the Paper I pilot study. The green fields labelled P and Q are additional MUSE-GTO observation for the outer spiral arm. The cyan fields indicate ML20 fields. The magenta fields are the previous PNe survey area of \protect\citet{2012A&A...547A..78P} using the FORS2 instrument. Image: NGC\,300 in \ha~taken with the Wide Field Imager (ESO) -- Program ID 065.N-0076}.
      \label{fig:fields}
   \end{figure*} 


\section{Observations and data reduction}

The data for this project were acquired using the MUSE spectrograph on the 8.2-meter Very Large Telescope \citep{2010SPIE.7735E..08B}. 9 fields were obtained as part of the MUSE guaranteed time observation (GTO) program\footnote{Program IDs 094.D-0116, 095.D-0173, 097.D-0348, and 0102.B-0317 -- PI: Roth}, while 35 fields were taken from the ESO Archive\footnote{Program ID 098.B-0193(A) -- PI: McLeod}. The area covered by these observations is shown in Figure \ref{fig:fields}. 

The initial MUSE-GTO data (field A, B, C, D, E, I, J) were obtained between the years 2014 -- 2016 using the extended wide field mode with a spatial coverage of $1' \times 1'$ and spectral coverage of $4650 - 9350$\,Å with $1.25$\,Å sampling. First results from the 7 fields in the centre area were reported in Paper I, covering the nucleus, part of the spiral arm that extends from the nucleus to the north-west, and inter-arm regions of the galaxy. In late 2018, additional fields P and Q were observed to cover the part of the outer spiral arm. Moreover, fields A, B, and C were re-observed with adaptive optics support to obtain better image quality. In the adaptive optics mode, a notch filter at $5750 - 6100$\,Å blocks the laser light, which is, however, not affecting the emission lines of interest. Most of the fields were obtained with an exposure time of $6 \times 900$ s, with the exception of field J ($4 \times 900$ s), field C ($8 \times 900$ s), and field B, D ($11 \times 900$ s). 

The 9 MUSE-GTO fields were reduced using the MUSE pipeline \citep{2020A&A...641A..28W} within the MUSE-WISE environment \citep{2015scop.confE..28V}, as explained in more detail in Paper III \citep{genovevangc300}. In addition to the field distortion correction produced by the pipeline, the astrometry is also calibrated using the Gaia DR2 catalogue \citep{2018yCat.1345....0G}, providing absolute astrometry within 0\farcs1.  Sky subtraction was performed using an offset field outside the galaxy. Since we will perform photometry in \oiii, we measure the seeing quality at this wavelength based on the FWHM of 3 to 4 stars for each field. These stars, which are typically giants or supergiants in the disk of NGC\,300, have apparent magnitudes of F606W $\gtrsim 21$ in the HST ACS magnitude system \citep{2018A&A...618A...3R}. Unlike the situation in more distant systems, such as Fornax cluster ellipticals \citep{2021ApJ...914...94S}, confusion with globular clusters is not a concern. For the MUSE GTO data, the image quality ranges from $0\farcs6 - 0\farcs8$ FWHM, as presented in Appendix \ref{appendix:fields}.

Another 35 fields, the ML20 data, publicly available at the ESO Archive, were originally observed to study stellar feedback in NGC\,300 \citep{2020ApJ...891...25M, 2021MNRAS.508.5425M}. The data were obtained using the nominal wide field mode, which has the same spatial coverage of $1' \times 1'$, but with a slightly shorter wavelength coverage of $4750 - 9350$\,Å. The observations were conducted in the period of 2016 -- 2018 without the support of adaptive optics, and each field was observed with an exposure time of $3 \times 900$s. These data were reduced using the fully automated MUSE pipeline \citep{2020A&A...641A..28W} with default parameters, as provided in the ESO Archive. The astrometry of the ML20 data only relied on the distortion correction within each field, which limited the absolute positional accuracy of the object catalogue to $\sim$ 3\arcsec. Moreover, the sky subtraction was performed using a reference region within each field instead of an offset field; this resulted in sky oversubtraction, especially in areas where diffuse gas is prominent. However, since we perform local sky subtraction for flux measurements of individual objects (see Section \ref{sec:analysis}), the effect cancels out. Based on our measurements, the seeing quality of the ML20 data in \oiii~ranges between $0\farcs8 - 1\farcs5$ FWHM\null.  These \oiii\ seeing measurements are presented in Appendix \ref{appendix:fields}.


\section{Data analysis}
\label{sec:analysis}

\subsection{PN detection and classification}
\label{subsec:class}

To find PN candidates, we employed the differential emission line filter (DELF) method described by \citet{2021arXiv210501982R}. This is performed by extracting 15 datacube layers around the wavelength of redshifted \oiii\ \citep[the systematic velocity of NGC\,300 is $v_\mathrm{{sys}} = 144 \: \mathrm{km/s}$;][]{1989spce.book.....L} and treating each layer as an on-band image; this 18.75\,Å range accounts for the different line-of-sight velocities (LOSV) within the galaxy. Then, an intermediate broadband continuum image is constructed from the wavelength range between $\lambda5063-5188$\,Å, which is free from strong absorption line features; this is used as the off-band image. By subtracting the scaled off-band image \citep[see scaling factor in Equation 8,][]{2021arXiv210501982R} from the on-band images, we obtain a series of continuum-free differential images. Using the DS9 software \citep{2003ASPC..295..489J}, the differential images are visually inspected to find the \oiii~sources. After experimenting unsuccessfully with DAOPHOT FIND \citep{1987PASP...99..191S} to identify PN candidates, which turned out to be unable to cope with the spatially variable emission line background in \oiii, we resorted to the datacube layer blinking technique, that is described in \citet{2021arXiv210501982R}.

The typical physical size of planetary nebulae is of the of order $\sim 0.3$ pc \citep{2006agna.book.....O}. If we assume a distance of 1.88 Mpc \citep{2005ApJ...628..695G} and a scale of $\sim 9 \: \mathrm{pc}/\arcsec$, we expect the PNe in NGC\,300 to appear as point sources. After marking the coordinates of the point sources in \oiii, we apply aperture photometry \citep{1987PASP...99..191S} at each wavelength layer along the datacube for these objects to obtain their spectra. We employ an aperture diameter of 3 spaxels (0\farcs6), an inner sky annulus of 12 spaxels, and an outer sky annulus of 15 spaxels. Although the seeing conditions of the datacubes vary between $\sim 0\farcs6 - 1\farcs5$, we extract the spectra using the same parameters. The small aperture of 3 spaxels is chosen to minimise contamination of background gas or nearby \hii~regions. Then, the line fluxes are extracted using Gaussian fitting with the LMFIT routine in Python \citep{newville2016lmfit}, keeping in mind that the MUSE data has a wavelength sampling of 1.25~Å\null. Since the MUSE-GTO and the ML20 data have overlapping areas, the classifications were done independently for each data set. Cross-matching was performed after the PNe were identified. 

To classify the sources into PNe, \hii~regions, and supernova remnants (SNR), we employed the BPT-diagram \citep{1981PASP...93....5B} that is based on the line ratio of \oiii/\hb~and \siisum/\ha. Besides the application of classifying active galaxies \citep{2001ApJ...556..121K, 2006MNRAS.372..961K}, the BPT-diagram has been demonstrated to effectively discriminate the PNe from their mimics \citep{2008MNRAS.384.1045K, 2010PASA...27..129F, 2013MNRAS.431..279S, 2021arXiv210501982R}. As the classification rely on emission line ratios with very similar wavelengths, we can assume that the relative line fluxes have negligible extinction and seeing dependency on wavelength.

\begin{figure}
   \centering
   \includegraphics[width=\hsize]{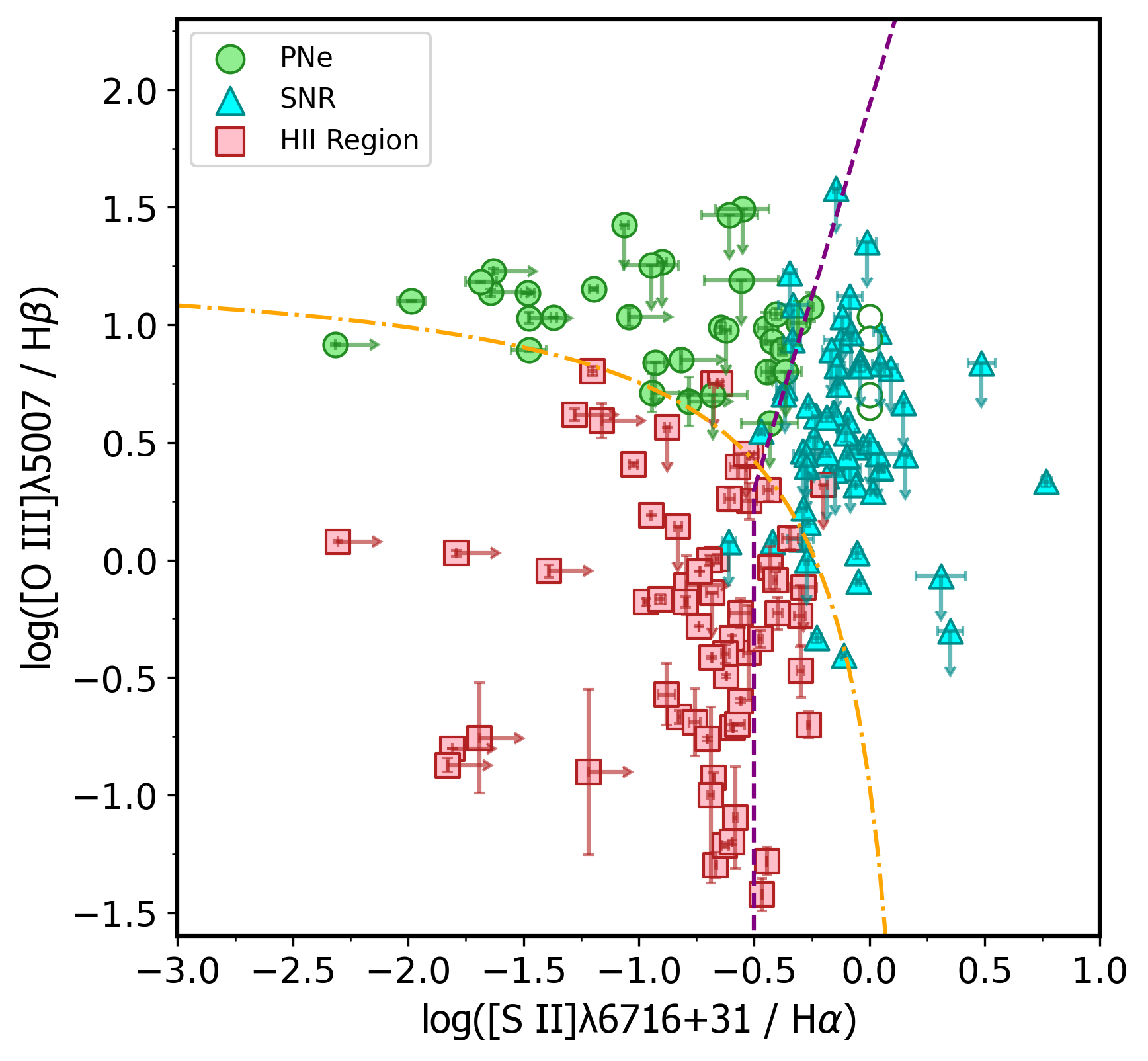}
   \includegraphics[width=\hsize]{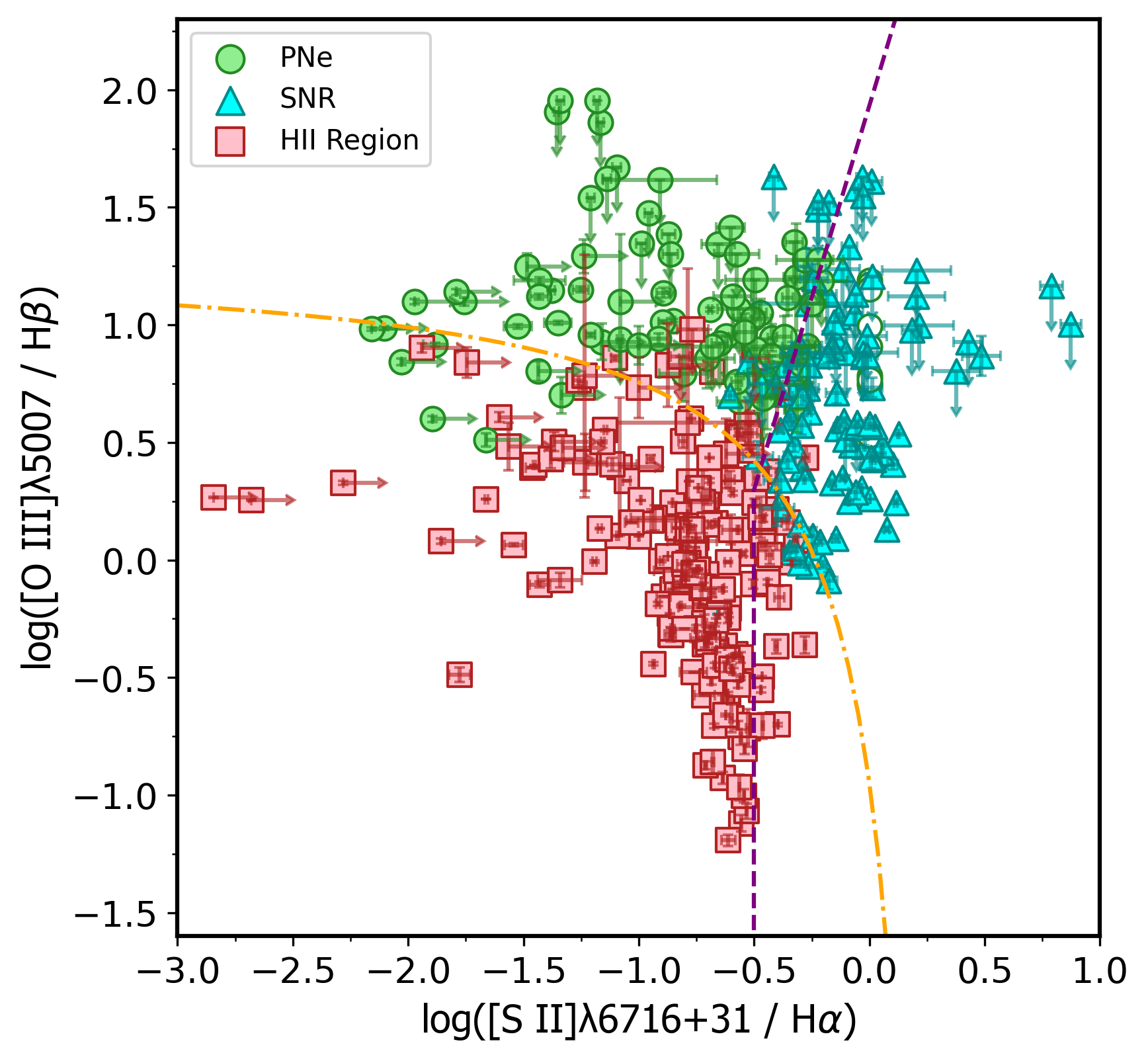}
      \caption{BPT-diagram of MUSE-GTO data (upper) and ML20 data (lower). The orange dot-dashed line is taken from \protect\citet{2001ApJ...556..121K} and the purple dashed line is defined by \protect\citet{2021arXiv210501982R}. Open circles indicate PN candidates, which only have \oiii~as the diagnostic line for the diagram. The deviation from the separation lines is explained in the text.}
      \label{fig:bpt-diagram}
   \end{figure}

Our BPT-diagrams for the MUSE-GTO and ML20 data are presented in Figure \ref{fig:bpt-diagram}. To separate the SNRs, we adopt the value log \siisum/\ha~$\geq -0.5$ from \citet{2021arXiv210501982R}. To discriminate PNe from \hii~regions, we employ the theoretical line by \citet{2001ApJ...556..121K}, which was originally intended to differentiate starburst galaxies. We also consider the line ratio of \siirat~as a proxy for density, since bright PNe are expected to be denser than both \hii~regions and SNRs \citep{2006agna.book.....O}. While the brightest \oiii~sources have sufficient line fluxes for the BPT-diagram classification, fainter sources may lack the weaker emission lines, i.e. the \hb~line or the \siidoublet~lines. In such cases, we assume lower limits for the line fluxes and classify the sources as PNe if the \oiii~line is stronger than the \ha~line, which may introduce deviation from the separation lines in the diagram. We also found some faint objects, which only have the detection of the \oiii~and the \nii~line without \ha~detection, which are possibly Type I PNe \citep{2010PASA...27..129F}. Moreover, we also put remarks for PNe, which are only classified solely based on \oiii~detection. This is the case for few of our faintest PN candidates. Nevertheless, such cases will not affect the distance determination because the PNLF cut-off is only defined by the brightest PNe.



In the MUSE-GTO data, we classified 37 PNe, 62 \hii~regions, and 59 SNRs. In ML20 data, we classified 85 PNe, 176 \hii~regions, and 105 SNRs. To cross-match the PNe candidates in the overlap area between the two dataset, we attempted an automated algorithm by comparing the sky coordinates. However, since the astrometric accuracy of both data differs, our attempt was not successful. Therefore, the cross-matching was performed visually using the DS9 software \citep{2003ASPC..295..489J}. The final PN number from the MUSE-GTO and ML20 fields: 105 PNe in the central region, and 2 in the P and Q fields at higher galactocentric distance. The PN catalogue is presented in Appendix \ref{appendix:catalogue}.


\subsection{PN number comparison}

Previous PN surveys of NGC\,300 were conducted by \citet{1996A&A...306....9S} -- SO96, \citet{2012A&A...547A..78P} -- PE12, and \citet{2018A&A...618A...3R} -- Paper I, who identified 34, 104, and 45 PN candidates, respectively. To demonstrate the accuracy of our classification, we employed the sample by PE12 as comparison, since it covers more area and contains more PNe than the other studies. PE12 observed NGC\,300 with the FORS2 imager \citep{1998Msngr..94....1A} in two $6.8' \times 6.8'$ fields, one in the centre, and another in the eastern outskirts of the galaxy. The study employed the on/off-band technique to detect PN candidates and classified objects based on the criterion of whether or not a central star was present in their 5105\,Å image. The expectation was that central stars of PNe would be too faint to be detected in the visual, while the ionising O stars in \hii\ regions could be seen. For brighter candidates with $m_{5007} < 25$, they also performed additional spectroscopy using the MXU-mode with the same instrument \citep{2013A&A...552A..12S}. Since our observations cover a smaller area on the sky, we only made the comparison for the intersecting $5' \times 7'$ region in the centre of the galaxy, which is also indicated in Figure \ref{fig:fields}.


In the overlapping region at the centre of the galaxy, we identify 105 PNe, compared to 58 in the PE12 sample. Moreover, although we recover all 58 sources found by PE12, our classification indicates several discrepancies.  While 43 of the PE12 sources are confirmed as PNe, we classify 9 objects as compact \hii\ regions and 3 as SNRs. These misclassifications could have happened due to the fact that PE12 only have the spectral classification for candidates with $m_{5007} < 25$, while the fainter objects completely relied on the detection of a central star. This approach also lacked the ability to identify SNRs amongst the fainter candidates, as such objects can be discriminated through the detection of the \siidoublet~lines \citep{2010PASA...27..129F, 2013MNRAS.431..279S}. Since all of our candidates are classified on the basis of their spectral properties, we believe that our classification is more reliable. Moreover, in terms of the number of detection, we also demonstrate that the MUSE observations are more sensitive and able to reach fainter magnitudes. 

\subsection{[O III]$\lambda$5007 photometry}
\label{section-photometry}

The \oiii~fluxes were obtained using DAOPHOT aperture photometry \citep{1987PASP...99..191S}, applied to the PNe candidates in the 15 differential layers for each datacube. Then, the magnitudes were computed using the $V$-band equivalent conversion \citep{1989ApJ...339...39J} defined as
\begin{equation}
    m_{5007} = -2.5 \: \mathrm{log} \: F_{5007} - 13.74
\end{equation}
where the flux is in erg~cm$^{-2}$~s$^{-1}$. Here, the aperture radius was adjusted to a value of approximately the FWHM of the PSF in a given exposure to accommodate the respective seeing condition. The inner and outer sky annulus were fixed to 12 and 15 spaxels, respectively. 
Most of the~flux of the PSF was obtained by adding the 5 bins closest to the Gaussian peak and the remaining flux is recovered through the use of an aperture correction based on the information of a PSF reference. This correction is crucial to obtain accurate fluxes, which however can be a challenge when there is no reference available especially with the small field of view of MUSE. The aperture correction method is explained in Appendix \ref{appendix:apcor}.

The photometric uncertainty was calculated from the Gaussian fit errors, convoluted with an assumed flux calibration error of 5\% \citep{2020A&A...641A..28W}. In high surface brightness regions of distant galaxies, double-peaked profiles are occasionally found and indicate the presence of two superposed PNe with different radial velocities \citep{2021arXiv210501982R}. Unsurprisingly, we do not find such cases in our sample. Since NGC\,300 is a quite nearby galaxy, spatial coincidences are less likely. Additionally, the five datacube layers containing the total flux for \oiii\  were also inspected to insure the PN candidate was not extended or contaminated by surrounding gas emission. 

\begin{figure}
   \centering
   \includegraphics[width=\hsize]{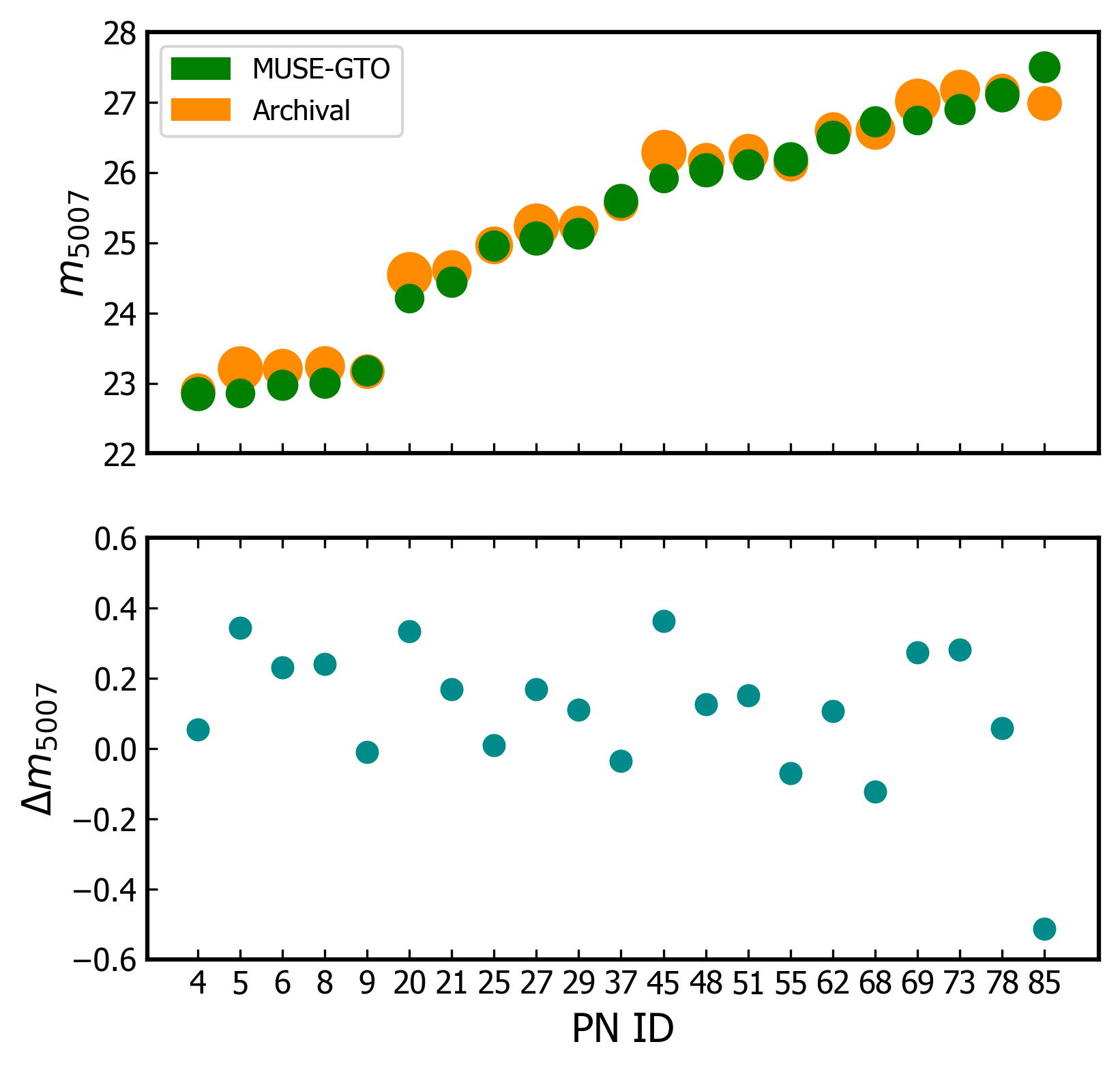}
   \includegraphics[width=\hsize]{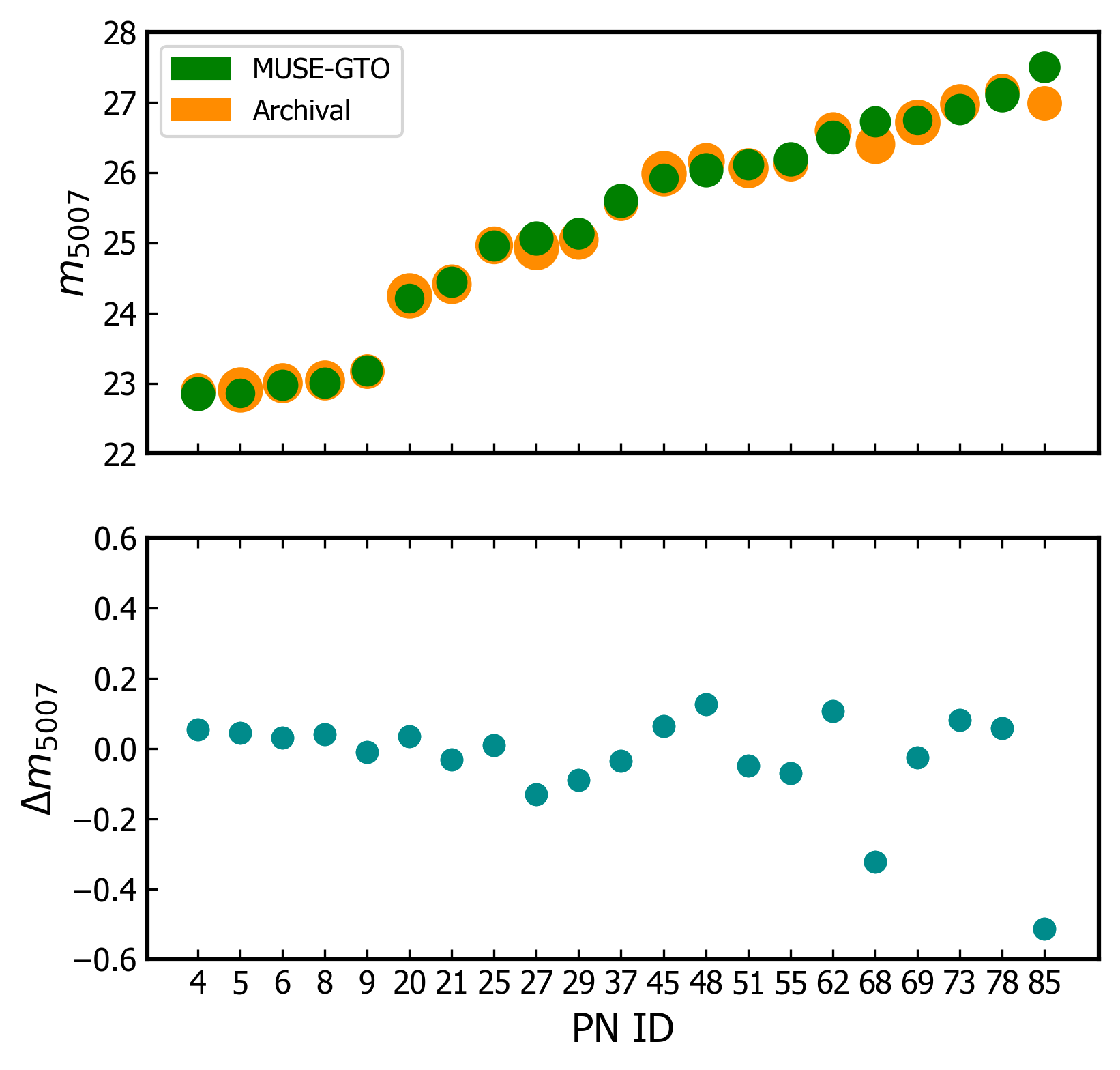}
      \caption{Comparison between the MUSE-GTO and ML20 \oiii\ magnitudes before (upper) and after (lower) photometric correction for PNe in the overlapping area. The markers are linearly scaled with the seeing FWHM of the field and sorted from the brightest to the faintest.}
         \label{fig:gto_vs_archive}
   \end{figure}

As an internal test of our photometry, we used the regions of field overlap to compare our PN measurements made in the MUSE-GTO fields to those from the ML20 data.   This comparison is presented in the upper panels of Figure \ref{fig:gto_vs_archive}. We find that the ML20 observations obtained under poor seeing condition tend to be systematically fainter than the MUSE-GTO data, while the photometry of the same object from different datacubes with similar image quality gives identical results (exception for the faintest PN in the comparison). Thus, the difference in seeing conditions can introduce a magnitude error; this is most likely due to the choice of a too small aperture for the asymptotic assumption for the aperture correction. Because the disk of NGC\,300 contains a large amount of diffuse emission-line gas, we chose to not to increase this radius. However, in order to obtain the same photometric quality between the two data sets, we applied and additional corrections of $0.2$ mag for objects with seeing FWHM $\sim 1\farcs2$ ($\sim 6$ spaxels), and $0.3$ mag for seeing FWHM $\sim 1\farcs4$ ($\sim 7$ spaxels). We found these values based on empirical trial and error to achieve the minimum average discrepancy between the two sets of magnitudes. 

The comparison after applying the correction is presented in the lower panels of Figure \ref{fig:gto_vs_archive}. The average discrepancy is now $0.05$ mag, which is still within the typical measurement error of $0.06$ mag. For PNe with $m_{5007} \sim 27$ or fainter, the correction has no meaningful implication, because the candidates are close to the detection limit. In the pilot study, \citet{2018A&A...618A...3R} performed a completeness simulation for the MUSE-GTO data of given exposure time, with seeing quality ranging from $0\farcs6 - 1\farcs2$. This was conducted by embedding artificial PNe with different magnitudes into the real datacubes. It was found that for a seeing of $1\farcs2$, the expected completeness is $90\%$ at $m_{5007} = 27$. Although we are able to detect a PN as faint as $m_{5007} = 28.91$, the seeing quality on average for all our data is $1\farcs0$, with $23\%$ of the fields exhibiting larger than $1\farcs2$ FWHM. This shows that completeness of $90\%$ at $m_{5007} = 27$ is only achieved for $77\%$ of our fields. However, since the emphasis of this work is on the bright candidates that define the PNLF cut-off for the distance determination, our results are not suffering from sample incompleteness at the faint end. For the final \oiii~magnitudes, we preferred the MUSE-GTO data, if available, and otherwise we employed the ML20 data. We also applied the correction for fields outside the overlapping area with seeing FWHM $> 1\farcs2$. In total, 11 PNe from 4 fields were corrected in this manner. 

\begin{figure}
   \centering
   \includegraphics[width=\hsize]{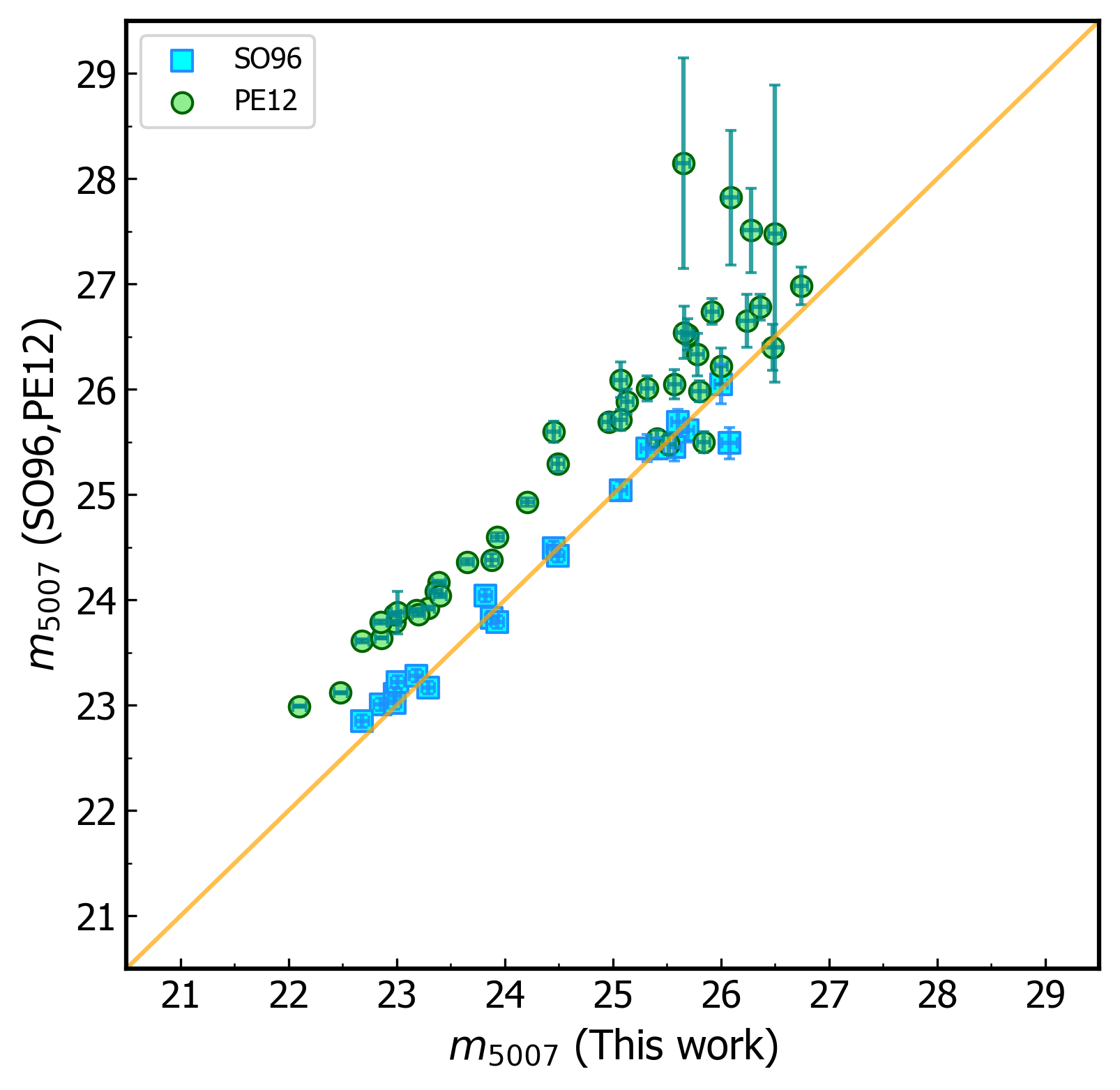}
      \caption{Comparison of $m_{5007}$ between this work, SO96, and PE12. The photometry of SO96 agrees within 0.01 mag. The photometry of PE12 is systematically fainter by 0.71 mag. For $m_{5007} > 25$, the relation with PE12 becomes scattered.}
         \label{fig:comparison_pena}
   \end{figure}

To test the accuracy of our photometry, we compared our magnitudes to the results from the literature. Figure \ref{fig:comparison_pena} shows a comparison with SO96 and PE12. While our data is in reasonable agreement with SO96 within 0.01 mag on average, there is a systematic offset with regard to PE12. We find that our magnitudes are systematically brighter by an average of 0.71 mag. PE12 obtained instrumental \oiii\ magnitudes for the FORS2 on-band image using aperture photometry with the aperture diameter of 5 pixels (1\farcs25), based on the average PSF FWHM of 2.9 pixels. To obtain the apparent $m_{5007}$ magnitudes, they calibrated the instrumental measurements using an empirical relation derived from the objects' spectroscopic fluxes, which are only available for the brightest PNe in their sample. 
We can try to understand what may be the reason for the discrepancy. First of all, we note that flux calibration is an established MUSE procedure in operation at the VLT and part of the data reduction pipeline. According to \citet{2020A&A...641A..28W}, flux calibration has been measured to be accurate to within 3-5\%. If a significant number of our MUSE exposures would have been affected by non-photometric observing conditions -- for which we have no evidence, we would expect a scattered, but not the tightly constrained linear correlation, that we see in Figure \ref{fig:comparison_pena}, in particular  for the brightest 3 magnitudes. Secondly, \citet{2018A&A...618A...3R} have tested synthetic MUSE datacube broadband photometry of stars against published HST photometry for the same GTO datacube subset that has been used in our work, showing no hint of an offset to within a magnitude of F606W=22.7. Thirdly, the agreement with SO96, who obtained their data with narrow-band imaging at the ESO NTT, i.e. a different instrument at a different telescope, gives us reasonable confidence that our flux calibration cannot be off by as much as a factor of almost 2. Finally, we can follow the argument put forward by \citet{2004ApJ...603..531R}, that by definition, integral field spectroscopy is an ideal tool for spectrophotometry, as is does not suffer from any kind of slit effects.

We can speculate, though, that the spectrophotometry from PE12 might have been affected in various ways to give rise to the observed systematic offset. In case of PE12, the calibration relies on spectroscopic fluxes, which were obtained using a slit spectrograph, and thus may suffered from slit-losses that were estimated to be $10 - 15\%$. However, from our comparison, we infer that the loss might be underestimated since 0.71 mag difference is equivalent to a loss of $\sim 48\%$. To test this, we performed a slit-loss simulation based on a model of the PSF with the quoted seeing conditions of $0\farcs7 - 0\farcs9$, and a slit size of 1\arcsec. The simulation is done in the $R$-band, as the seeing measurement is typically done in this band. Based on these parameters, our simulation predicts that the slit-losses should be between $8 - 20\%$. However, since the PSF FWHM is expected to be larger in the blue wavelength region, the slit-loss in \oiii~will be larger than that in the $R$-band. Moreover, additional losses can be introduced by positioning and guiding errors, as investigated by \citet{1993ApJ...417..209J}. Spectrophotometry with a slit spectrograph also requires the slit to be oriented at the parallactic angle to minimise the effect of atmospheric dispersion \citep{1982PASP...94..715F, 1993ApJ...417..209J}. Since our observations are performed with an IFU, we are not affected by any of these problems.  While it is possible that our use of small apertures has caused some flux to be lost, we are able to compensate for this loss using aperture corrections as described above. Such a procedure is not easily performed for data observed with a slit spectrograph. 

Besides the issue of slit-losses, the follow-up spectroscopy by PE12 is limited to PNe with $m_{5007} < 25$, which is less than $~40\%$ of their whole sample. This implies that most of their PNe are dependant on the measurement accuracy of the brighter PNe, which are likely to be affected by systematic errors. Moreover, the use a 5 pixel aperture to measure the flux for a 2.9 pixel FWHM PSF incurs a risk of including light from background contamination. In such crowded fields with a variable background and ubiquitous diffuse emission-line gas, the aperture might unexpectedly collect \oiii\ flux of the ambient interstellar medium. Without proper background inspection and subtraction, this might lead to an overestimation of brightness. The inclusion of background emission likely explains the scatter for $m_{5007} > 25$ in Figure \ref{fig:comparison_pena}. The spatial resolution of MUSE allows us to carefully check and analyse the condition of the background on a case-by-case basis and provide more accurate photometry. The variable background is also the main consideration to opt for a smaller aperture size for our flux measurements, and to rely on the aperture correction to deliver the final values.

\subsection{Balmer decrement}

Measurement of extinction using the Balmer decrement with \ha/\hb~ratio have been demonstrated on MUSE data for different objects, i.e. Pillars of Creation in M16 \citep{2015MNRAS.450.1057M}, core of R136 in the LMC \citep{2018A&A...614A.147C}, faint \hii~regions in NGC\,300 \citep{genovevangc300}. To obtain this, we employed the spectra extracted for the classification, as explained in Section \ref{subsec:class}, using the aperture of 3 spaxels, with the inner and outer sky annulus of 12 and 15 spaxels, respectively. We also apply aperture correction for the Balmer lines, which can be referred to in Appendix \ref{appendix:apcor}.

However, not all of our PNe candidates are detected at these two wavelengths. In order to filter out the candidates, we put a threshold of $F(\mathrm{H}\alpha) = 2 \times 10^{-17}$ erg~cm$^{-2}$~s$^{-1}$. For typical PNe with electron temperature $T_e = 10.000$ K, the expected Balmer ratio is $\mathrm{H}\alpha / \mathrm{H}\beta= 2.86$ \citep{2006agna.book.....O}, which corresponds to $F(\mathrm{H}\beta) \sim 8.75 \times 10^{-18}$ erg~cm$^{-2}$~s$^{-1}$ for the \ha~threshold. Any \hb~flux lower than the threshold is too close to the detection limit. For such cases, we assume the upper limit of \hb~flux derived from the \ha~line, which consequently also assumes no extinction. To avoid possible biased exclusion of high extinction PNe, we flag the objects with upper limit \hb~flux. If the \ha~flux is below the threshold, then the extinction measurement is not performed. Based on the \ha~threshold criterion, we have a complete sample for PNe down to $m_{5007} = 24.5$. If we extend the sample to fainter magnitudes, we reach $87\%$ completeness until $m_{5007} = 26.0$ and $64\%$ completeness until $m_{5007} = 27.0$. To calculate the extinction, we then used the Balmer decrement defined as


\begin{equation}
    A_\lambda = k(\lambda) \: c(\mathrm{H}\beta) = \frac{ k(\lambda) \: 2.5}{k(\mathrm{H}\beta)-k(\mathrm{H}\alpha)}\bigg{[}\mathrm{log} \: \bigg{(}\frac{\mathrm{H}\alpha}{\mathrm{H}\beta}\bigg{)} - \mathrm{log}\:(2.86) \bigg{]}
    \label{eq:extinction}
\end{equation}
where $k(\lambda)$ is the wavelength dependant extinction constant. For the foreground extinction, we employed the extinction curve of \citet{1989ApJ...345..245C} with $R_V = 3.1$ and $E(B-V) = 0.011$ \citep{2011ApJ...737..103S}. For NGC 300, \citet{2009ApJ...700..309B} measured the present day metallicity of $12 + \mathrm{log(O/H)} \sim 8.1 - 8.5$ using the \hii~regions. Recent measurement using the same MUSE-GTO data based on faint \hii~regions and diffuse interstellar gas (DIG) also agrees with the latter value as $12 + \mathrm{log(O/H)} \sim 8.5$ \citep{genovevangc300}. Since the chemical abundance of NGC 300 in the observation area are similar to the LMC, with $12 + \mathrm{log(O/H)} \sim 8.4 - 8.5$ \citep{2017MNRAS.467.3759T}, we employed the average LMC extinction curve to obtain the extinction of our PNe. The uncertainty of our extinction measurement is highly dependent to the aperture correction method. Therefore, we quote an estimated error based on
the comparison of extinction calculated from different aperture
correction methods, as explained in Appendix \ref{extinction_uncertainty}.

\begin{figure}
   \centering
   \includegraphics[width=\hsize]{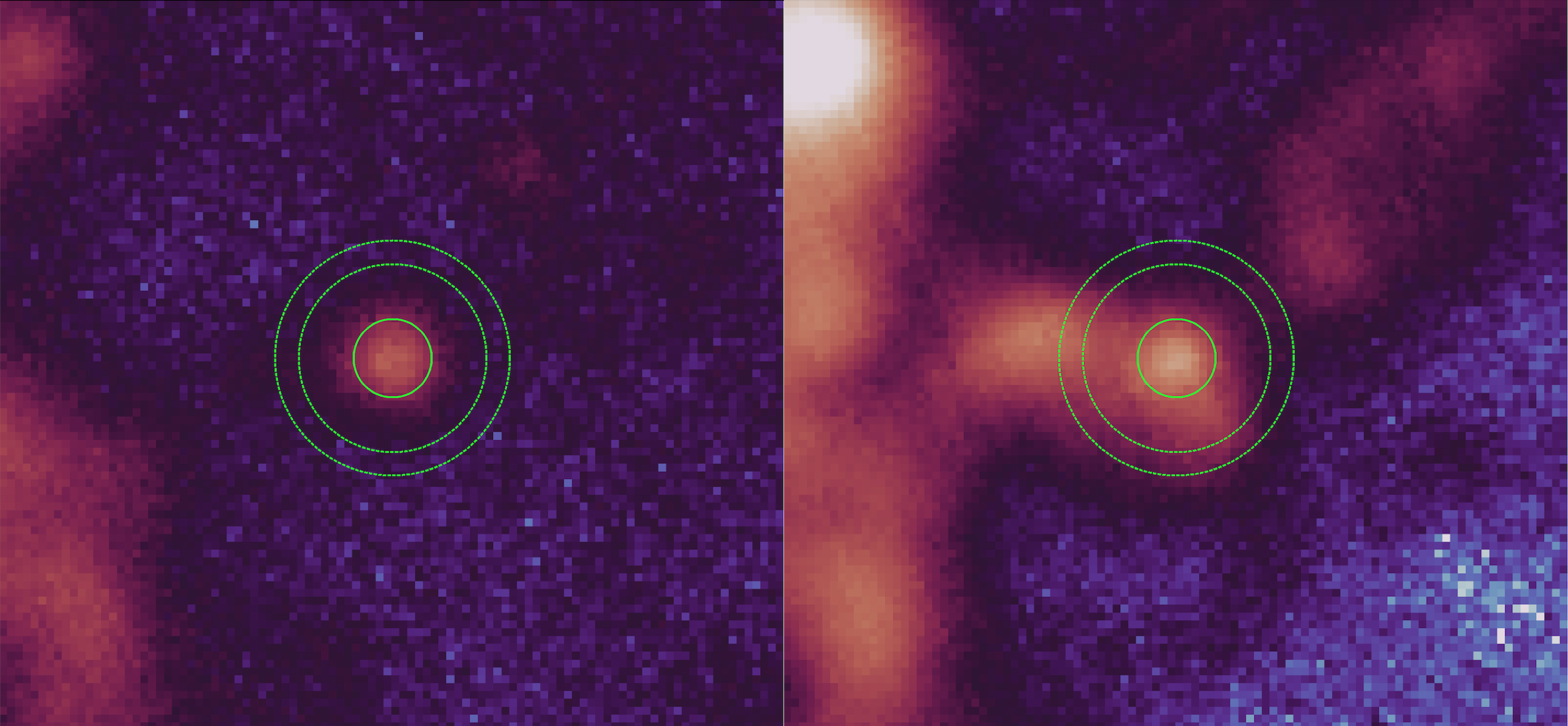}
  \caption{False-colour spatial map in \oiii~(left) and \ha~(right). The flux scaling is identical and logarithmic. The images are $20\arcsec \times 20\arcsec$ ($\sim 180 \times 180$ pc) each. The green marker illustrate the main and sky aperture. The candidate is isolated in \oiii, but overlapped with nearby \hii~region in \ha.}
         \label{fig:map}
   \end{figure}

In spiral galaxies, \citet{2009ApJ...703..894H} found that the typical extinction for the PNe in \oiii~is $A_{5007} \sim 0.7$. However, we discovered three high extinction cases with $A_{5007} > 1.5$, including one with an extreme value of $A_{5007} \sim 3.3$. While it is possible that some PNe exhibit high intrinsic extinction, as high metallicity populations and massive progenitors tend to produce dustier PNe \citep{2012ApJ...753..172S}, we suspect that the Balmer decrement might not always be accurate due to the local contamination. To investigate this further, and to highlight possible pitfalls that may play a role in studies based on slit spectroscopy, we examined spatial maps of the high-extinction objects in the wavelengths of \hb, \oiii, and \ha, using the p3d software \citep{2010A&A...515A..35S}. We found that these PNe candidates are co-spatial with nearby \hii~regions. In Figure \ref{fig:map}, a PN is shown to be an isolated point source in \oiii. However, in the spatial map of \ha, the point source is entirely embedded inside the extended emission surface brightness distribution of an unrelated nebula. This clearly shows that the Balmer line flux of the PN candidate is contaminated, and an accurate extinction measurement of the PN itself cannot be obtained. All three of the objects in question show similar patterns of contamination. We therefore excluded them from the sample. 

We compared our PN extinction measurements with the results from \citet{2013A&A...552A..12S} -- also referred as ST13 -- who observed PNe in NGC\,300 with the FORS2-MXU instrument at the VLT \citep{1998Msngr..94....1A}. They used 3 grisms 600B, 600RI, and 300T to cover spectral ranges of $3600 - 5100$Å, $5000 - 7500$\,Å, and $6500 - 9500$\,Å, respectively. To avoid uncertainties from the flux calibration of different bands, they employed the \hg~and \hb~lines from the 600B grism spectra to measure the Balmer decrement. The extinction values for 18 PNe in common are presented in Table \ref{st13_compare}. 

We have contemplated several reasons to explain the discrepancy. Firstly, slit losses that occur for the measurement of \oiii~most likely also occur for the \hg~and \hb~lines.
The short baseline between the lines is also very sensitive to systematic errors, making it difficult to derive accurate extinction values. Moreover, higher order Balmer lines are typically weak. The accuracy of measuring their flux depends on the precision to which the background of stellar absorption line spectra can be subtracted \citep{1993ApJ...417..209J, 2004ApJ...603..531R}. Using the PMAS instrument, \citet{2004ApJ...603..531R} compared the accuracy of the \hb~line flux obtained with the IFU and simulated slits. They demonstrated that the orientation of the slit can introduce a different sampling of the background, leading to systematic differences of the derived flux measurements. Since ST13 performed their measurements with slit spectroscopy, they were susceptible to this type of error. 

\begin{table}
\caption{\label{st13_compare} Extinction comparison between this work and ST13.}
\centering
\begin{tabular}{cccc}
\hline\hline
ID$_\mathrm{{MUSE}}$ & ID$_{\mathrm{ST13}}$ & c(H$\beta$)$_\mathrm{{MUSE}}$ & c(H$\beta$)$_\mathrm{{ST13}}$ \\
\hline \\[-3.2mm]
E-11 & 12 & 0.12$\pm$0.07 & 0.00 \\[1mm]
E-2 & 14 & 0.10$\pm$0.07 & 0.00 \\[1mm]
L6-5\tablefootmark{a} & 20 & -- & 1.17 \\[1mm]
L6-7\tablefootmark{a} & 22 & 0.20$\pm$0.09 & 0.44 \\[1mm]
I-2\tablefootmark{b} & 24 & 0.02$\pm$0.04 & 0.12 \\[1mm]
C-7\tablefootmark{b} & 25 & 0.03$\pm$0.10 & 0.07  \\[1mm]
H9-1 & 35 & 0.04$\pm$0.08 & 0.00 \\[1mm]
L9-8 & 40 & 0.13$\pm$0.07 & 0.00 \\[1mm]
H2-6 & 45 & 0.17$\pm$0.11 & 0.00 \\[1mm]
A-23 & 48 & 0.42$\pm$0.05 & 0.21 \\[1mm]
A-11 & 51 & 0.10$\pm$0.05 & 0.00 \\[1mm]
H1-8 & 54 & 0.19$\pm$0.06 & 0.00 \\[1mm]
H7-2 & 58 & 0.11$\pm$0.08 & 0.00 \\[1mm]
H1-1\tablefootmark{c} & 63 & -- & 0.41 \\[1mm]
H6-5 & 65 & 0.10$\pm$0.07 & 0.00 \\[1mm]
L2-3 & 66 & 0.05$\pm$0.08 & 0.00 \\[1mm]
H6-3 & 69 & 0.26$\pm$0.07 & 0.00  \\[1mm]
H12-1\tablefootmark{a} & 74 & -- & 0.64 \\
\hline
\end{tabular}
\tablefoot{\\
\tablefoottext{a}{severe Balmer contamination}\\
\tablefoottext{b}{uniform diffuse \ha~background}\\
\tablefoottext{c}{low excitation -- possibly compact \hii~region}}
\end{table}

For cases where the internal extinction of ST13 is reported higher than our values, we find that 3 of their PNe are discarded from our sample (L6-5, H1-1, and H12-1 in Table \ref{st13_compare}), either because of severe contamination as shown in Figure \ref{fig:map}, or by their low excitation, which we consider typical for compact \hii~regions \citep{2010PASA...27..129F}. We also found that in some of these cases, the PN is embedded in diffuse gas. In our sample, if the diffuse gas is assumed to be uniformly distributed, the flux excess can be corrected using the background sky annulus. It remains an open question whether the background correction of diffuse gas was accurately accounted for in the slit spectroscopy of ST13, but we conclude that a careful consideration of background subtraction is critical for the extinction measurements based on the Balmer decrement.


\section{The PNLF}
\label{section:pnlf}

\begin{figure}
   \centering
   \includegraphics[width=\hsize]{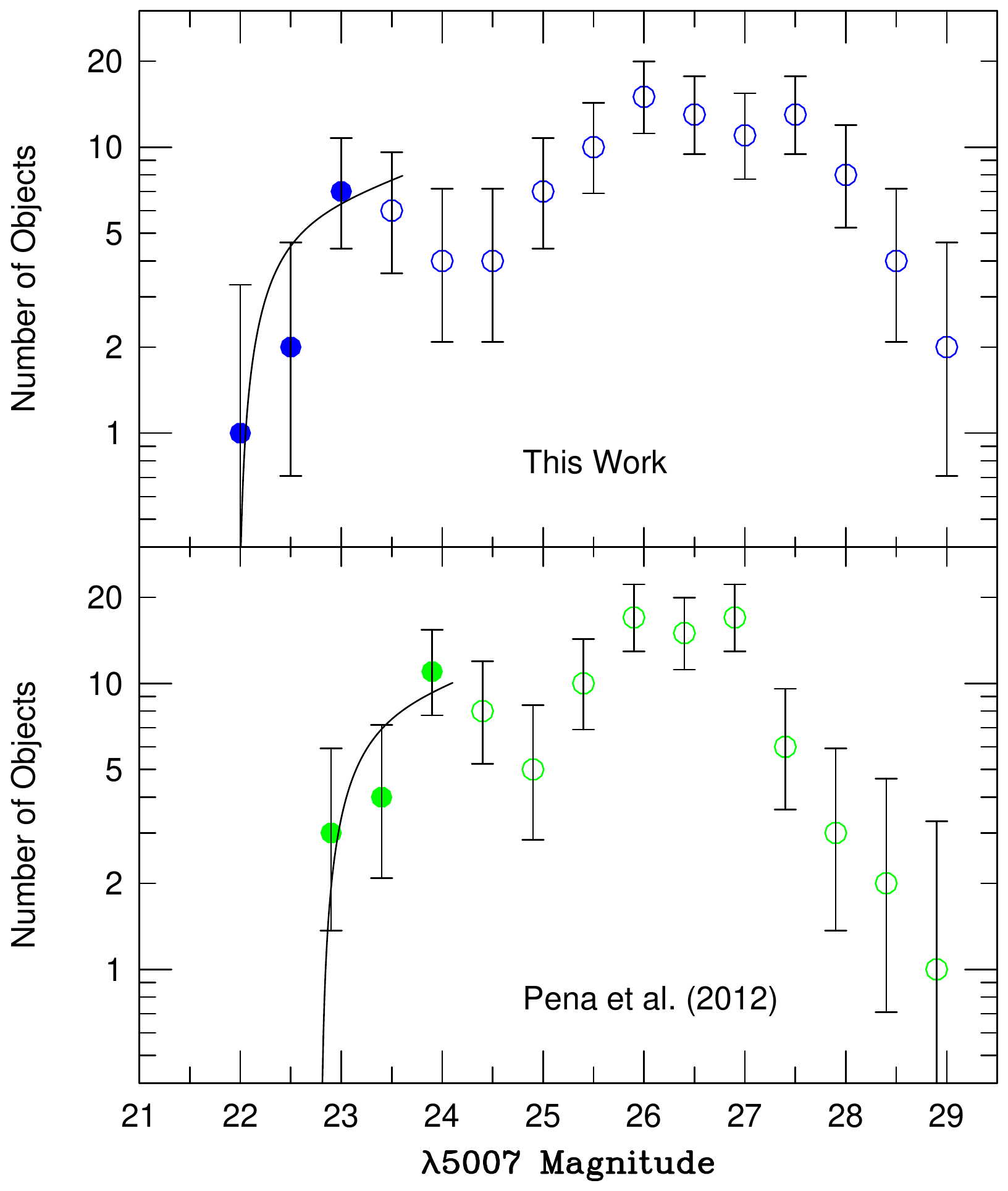}
      \caption{PNLF of NGC\,300 using MUSE (top) and PE12 photometry (bottom). Completeness limit for distance measurement is assumed at $m_{5007} = 23.6$ and $m_{5007} = 24.0$ for ours and PE12, respectively. Open symbols indicate incompleteness. The PNLF dip is visible for both.}
         \label{fig:pnlf}
   \end{figure}

The PN luminosity function of this work is presented in Figure \ref{fig:pnlf}. It exhibits the dip between 1 and 3 magnitudes below the cut-off. Such dip is typically observed in star-forming galaxies \citep{2002AJ....123..269J, 2010PASA...27..149C, 2010MNRAS.405.1349R}, which is possibly caused by multiple episodes of star formation \citep{2015A&A...575A...1R, 2021A&A...647A.130B} or difference in opacity and mass range of the PN formation \citep{2019AAS...23315001V}.


To determine the distance, we employed the maximum likelihood technique, where the empirical PNLF is treated as a probability function \citep{1989ApJ...339...53C}, assuming $M^*=-4.53 \pm 0.06$ and a fixed slope parameter of 0.307. When the number of PNe at the bright end cut-off is less than $\sim50$, distance determinations based on $\chi^2$ minimisation depend significantly on the details of how the PNLF is binned. Such methods are not recommended \citep{1989ApJ...339...53C, 2021arXiv210501982R}. Although our observation extends to $m_{5007} \sim 29$, the PNLF fit is only performed for the sample brighter than the dip until $m_{5007} = 23.6$, since equation~(\ref{eq:pnlf}) does not consider the dip feature, which nevertheless is insignificant for the distance determination \citep{2021A&A...653A.167S, 2022FrASS...9.6326C}. By taking the foreground extinction of $E(B-V) = 0.011$  \citep{2011ApJ...737..103S} into account, the most-likely distance modulus is $(m-M)_0 = 26.48^{+0.11}_{-0.26}$ with the uncertainties representing the statistical error of the fit and the $M^*$ uncertainty.

We also calculated the distance using PE12 photometry to make a comparison. Assuming a completeness limit of $m_{5007} = 24.0$, our maximum likelihood approach yields $(m-M)_0 =  27.30^{+0.09}_{-0.20}$, a value that is significantly larger than our MUSE distance.  We argue that this is due to the systematically fainter magnitudes of PE12 photometry, as discussed in Section \ref{section-photometry}. It is important to mention that PE12 measured a modulus distance of $(m-M)_0 = 26.29^{+0.12}_{-0.22}$, a value much smaller than our maximum likelihood values, including our distance measurement using the PE12 data. This difference is possibly due to their use of the Levenberg-Marquardt $\chi^2$ fitting technique, which is dependant on the binning method to construct the luminosity function. Since the  sample size is limited, they employed rather wide magnitude bins of 1.16 mag, which did result in a luminosity function shape that closely resembles the empirical law. However, in the luminosity function of PE12, their first magnitude bin is located at $m_{5007} \sim 22$, despite the fact that their brightest PN has a magnitude of $m_{5007} = 22.99$.  Thus, when the fit is performed, this systematic shift to brighter magnitudes results in a smaller distance modulus. This demonstrates that the choice of bin size can produce unintended systematical shifts of the luminosity function when the number of PNe in the top $\sim 0.5$~mag of the luminosity function is small.  Similarly, PE12's choice of bin size also smeared out detail in the PNLF's shape, as they did not report the observation of the PNLF dip. In Figure \ref{fig:pnlf}, we present the PNLF that we plot with the original data from PE12 using higher binning resolution than the original work.  In fact, the dip is present in the PE12 data, confirming that it is not an artefact in our measurements.

Finally, PE12 employed a larger extinction correction for the photometry with $A_{5007} = 0.2$ compared to our value of $A_{5007} = 0.05$. PE12 assumed this as the intermediate value between found by \citet{2005ApJ...628..695G} with $E(B-V) = 0.096$ ($A_{5007} = 0.3$) and \citet{1998ApJ...500..525S} with $E(B-V) = 0.013$ ($A_{5007} = 0.05$). In the case of \citet{2005ApJ...628..695G}, the extinction value is the sum of both the foreground extinction of \citet{1998ApJ...500..525S} and internal extinction derived from the Cepheids. For Cepheid distances, the internal extinction correction is necessary since they are originated from Population I stars, that are typically surrounded by galactic dust. However, this is less true for the PNe, so the foreground extinction correction for the PNLF distance is sufficient \citep{2010PASA...27..149C}. This implies that the extinction correction $A_{5007} = 0.2$ by PE12 is overestimated and also contributes to the smaller distance modulus. Therefore, the discrepancy between our calculation of the PE12 data and the original calculation is traced back to the issue of binning a limited sample and also the extinction correction. 

\section{Discussion}
\subsection{PNLF Distance}

\begin{figure*}
   \centering
   \includegraphics[width=\hsize]{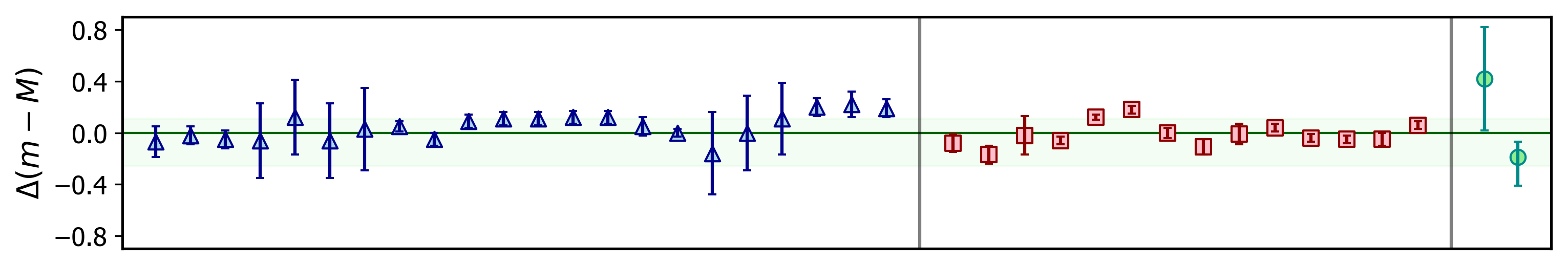}
      \caption{Distance modulus difference between our PNLF result with Cepheids (blue triangles) and TRGB (red squares), obtained from NED and sorted based on publication date. The Cepheid distances are from \citet{2001ApJ...548..564W, 2002A&A...389...19P, 2004AJ....128.1167G, 2005ApJ...628..695G, 2006ApJS..165..108S, 2010ApJ...715..277B,2016AJ....151...88B}. The TRGB distances are from \citet{2004AJ....127.1472B, 2004ApJ...608...42S, 2005A&A...431..127T, 2006AJ....132..729T, 2006ApJ...638..766R, 2007ApJ...661..815R, 2009AJ....138..332J, 2009ApJS..183...67D}. Previous PNLF distances of \citet{1996A&A...306....9S} and \citet{2012A&A...547A..78P} is also presented (green circles). The green shadow indicated the uncertainty of our PNLF distance.}
         \label{fig:cepheid_trgb}
   \end{figure*}

To demonstrate the accuracy of our distance measurement, we compare our result with previous distances in the literature derived using Cepheids and tip of the red giant branch (TRGB), taken from NED, in Figure \ref{fig:cepheid_trgb}. We can see that most of the distances are within the uncertainties of our PNLF result. One aspect that may introduce the discrepancy is the correction of extinction. For instance, the Cepheid distance of \citet{2005ApJ...628..695G} and the TRGB distance of \citet{2006ApJ...638..766R}, both part of the Auracaria project, are corrected with foreground and internal extinction of $E(B-V) = 0.096$. However, \citet{2007ApJ...661..815R} argue that the extinction derived from dusty young Cepheids by \citet{2005ApJ...628..695G} are not representative for the whole galaxy; their result only applies the foreground component of extinction. This shows the importance of having the same zero-point when comparing different distances derived from different methods. 

Moreover, we also show that the MUSE observation, combined with the differential emission line filter  \citep[DELF,][]{2021arXiv210501982R} and maximum likelihood technique \citep{1989ApJ...339...53C}, has improved the accuracy of PNLF method, as shown in Figure \ref{fig:cepheid_trgb}. The early result by \citet{1996A&A...306....9S} is based on the limited sample of only 34 PNe, from which they only construct a cumulative PNLF and employed the distance modulus of the LMC as a yardstick.  \citet{2012A&A...547A..78P} identified a significantly larger sample of 104 PNe, but as shown in Section \ref{section-photometry}, their data may suffer from slit-losses and contamination. The systematically fainter PN magnitudes then led to larger distance modulus, as described in Section \ref{section:pnlf}. Since the cut-off of the PNLF of NGC\,300 is defined by a very small number of PNe, minimisation fitting methods become too dependant on the binning \citep{1989ApJ...339...53C}. In a study by \citet{1997eds..proc..197J}, a correction for PNLF distance based on the number of PN sample is suggested. For a PNLF cut-off sample $< 20$ PNe, they estimated a distance correction of $\sim 0.1$ mag (see Figure 5 in \citealp{1997eds..proc..197J}). However, since the Cepheid and TRGB distance also varies with standard deviation of $0.1$ mag, there are no solid distance reference to test if the correction is appropriate. Nevertheless, we have shown that the PNLF distance derived with 
the maximum likelihood technique is more robust. We take this as a motivation to improve PNLF distance measurements for nearby galaxies with our method.

\subsection{Local dust effect on PN extinction}

Dust formation plays important role in the early stages of PN evolution since it occurs at the surface of the progenitor AGB star and presumably plays an important role in the envelope ejection \citep{2005ARA&A..43..435H, 2012ApJ...753..172S}. Infrared studies in the Milky Way and the LMC have revealed that the dust production is dependant on metallicity, with dustier systems found in higher metallicity environments \citep{2007ApJ...671.1669S,2012ApJ...753..172S, 2009ApJ...699.1541B}. Although it cannot tell the properties of the dust, Balmer decrement extinction measurements can also probe the presence of dust in PNe. In the study of \citet{2018ApJ...863..189D}, a comparison was made between the PN extinction distribution in the bulge of M31 and several other galaxies: the LMC \citep{2010MNRAS.405.1349R}, NGC\,4697 \citep{2008ApJS..175..522M}, and NGC\,5128 \citep{2012A&A...544A..70W}. Despite the limited samples involved, the authors found that the average extinction of PNe in each galaxy roughly follows the metallicity of the system. 

  \begin{figure}
   \centering
   \includegraphics[width=\hsize]{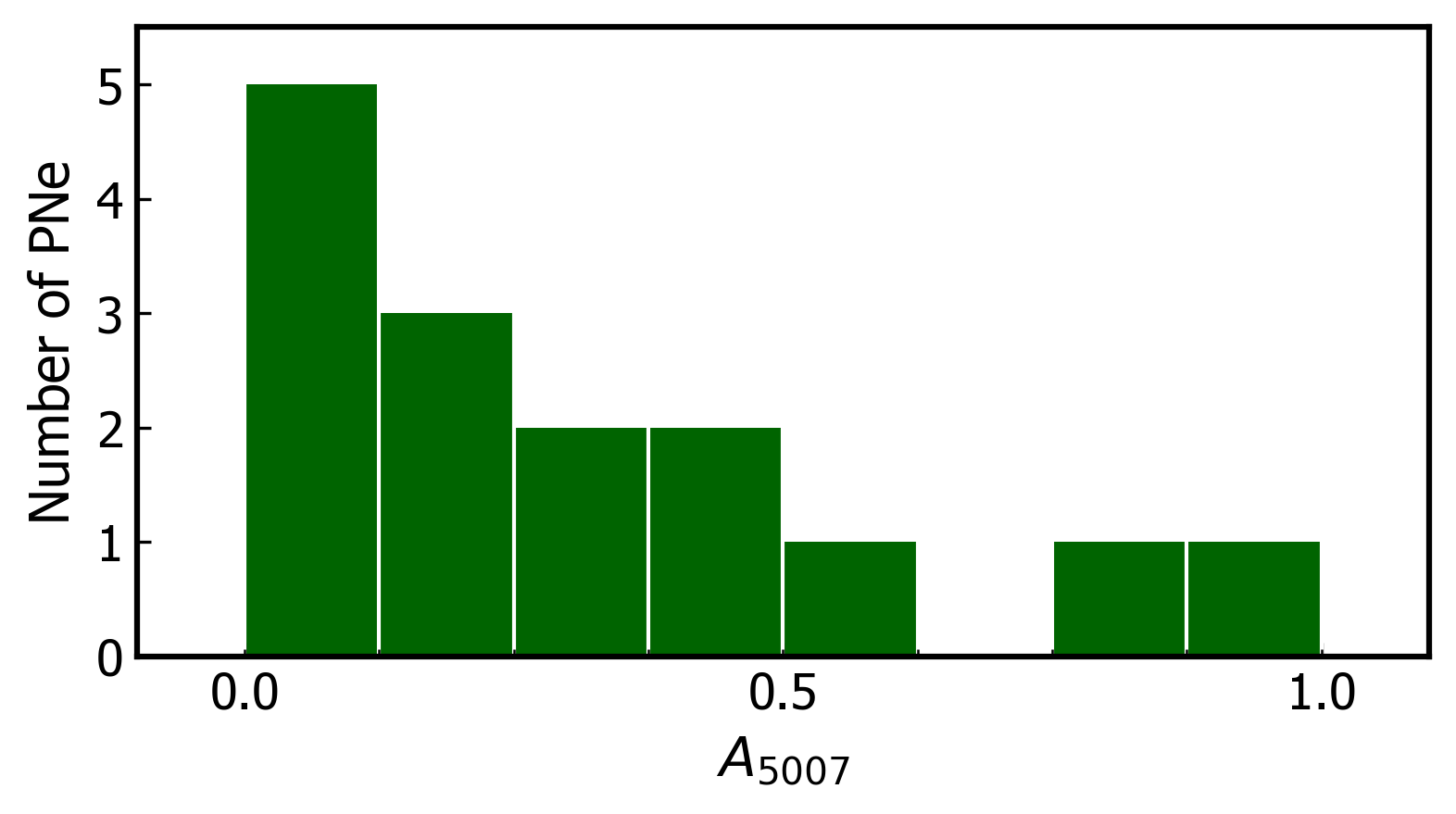}
      \caption{Distribution of extinction measurement for the PNe in \oiii~until $m_{5007} = 23.6$. The average extinction is $A_{5007} = 0.31$ ($c(\mathrm{H}\beta) \sim 0.09$).}
         \label{fig:extinction_distribution}
   \end{figure}

To investigate such trends in NGC\,300, we plot the extinction distribution in \oiii~for our PNe until $m_{5007} = 23.6$ (15 PNe) in Figure \ref{fig:extinction_distribution}. These PNe are the ones employed for the maximum likelihood distance measurement. We find that in general these bright PNe have low extinction in \oiii. The average extinction value for this sample is $A_{5007} = 0.31$ ($c(\mathrm{H}\beta) \sim 0.09$), which is lower than the average of the bright PNe sample in the LMC with $A_{5007} = 0.57$ \citep{2010MNRAS.405.1349R, 2018ApJ...863..189D}. However, we refrain from further interpreting the extinction distribution with the PN dust production, since the distribution is likely to be affected by local dust clouds, which can vary from one object to another. Such a problem has been reported in NGC\,5128, where the high extinction of some PNe was attributed to local dust clouds rather than the PNe themselves \citep{2012A&A...544A..70W}.  

    \begin{figure}
    \centering
    \includegraphics[width=\hsize]{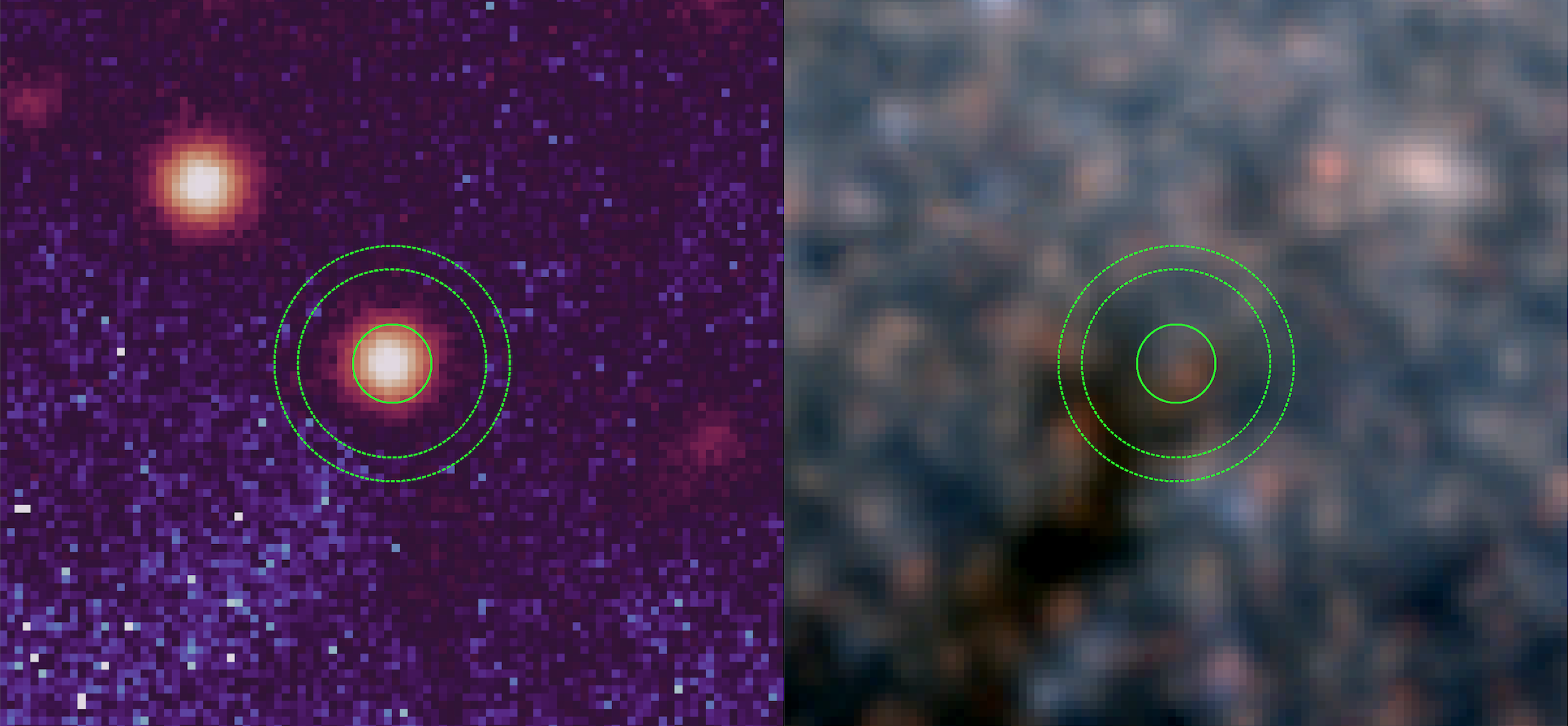}
      \caption{False colour (left) and RGB colour (right) map of the region surrounding PN A-23. The flux scaling is logarithmic. The images are $20\arcsec \times 20\arcsec$ ($\sim 180 \times 180$ pc) each. The green marker illustrate the main and sky aperture. The dust patch is clearly visible in the RGB image, overlapped with PN A-23.}
         \label{fig:dust-map}
    \end{figure}
    
At the distance of NGC\,300, our MUSE observations offer a spatial resolution of between 6 and 14 pc.  This resolution should be sufficient to visually resolve the spatial variation of dust extinction \citep{2013ApJ...771...62K, 2017ApJ...844..155T}. To test this, we inspected several objects with high extinction. As an example, we present the spatial map in \oiii~and RGB colours, which is constructed from Johnson-VRI filters, for the PN with the highest extinction value (PN A-23, $A_{5007} = 1.18$) in Figure \ref{fig:dust-map}. Although it is not obvious in the \oiii~image, the RGB image shows a dust lane patch, extending from the lower left corner to the centre. Since PN A-23 is in proximity to the dust lane, we suggest that the measured high extinction of this object is composed of both local dust within the galaxy and circumnebular extinction associated with the PN itself. 

Based on the comparison study between Balmer decrement extinction and infrared dust distribution in M31, \citet{2017ApJ...844..155T} concluded that vertical distribution of diffuse interstellar gas (DIG) and dust can vary in different locations of the galaxy and thus cause differing amounts of extinction. For NGC\,300, variation of extinction also has been reported by \citet{2005ApJ...632..227R}. Therefore, there is currently no guarantee that the measured extinction of individual PNe is free from local effects, which is confirmed with our images. We must therefore refrain from making conclusions based on the extinction values alone, until the different components of the extinction can be quantitatively resolved.

Although the extinction of individual PNe might be affected by local dust lanes, such effects are less significant for the luminosity function as a whole. The effect of dust scale height in the PNLF distances of late-type disk galaxies has been discussed by \citet{1997ApJ...479..231F}. They modelled PNLF with varying extinction in \oiii~and concluded that the inferred distance modulus should always be within 0.1 mag of the derived distance without extinction. A similar result also obtained by \citet{2005MNRAS.361..330R}, who modelled the PNLF with different scale heights of dust in the starburst galaxy NGC\,253. They found that even when the disk was optically thick with 1~mag of extinction, the PNLF distance is robust to within 0.1~mag.  Both studies suggest that the brighter PNe tend to be located above the dust layer from the point of view of the observer, or for other reasons suffer little extinction from within the galaxy. With these arguments, we do not expect the occurrence of dust lane extinction to significantly affect our distance result and a correction for internal extinction is at this point not necessary. 

\subsection{PN parent populations}
\label{section:parameter}

To gain a better understanding of the parent population of the PNe, we estimate the luminosity and the effective temperature of the central stars of the planetary nebula (CSPNs). These parameters are calculated for PNe until $m_{5007} = 26$ with measurable extinction, which corresponds to $87\%$ of the objects within this magnitude limit.

Simulation studies suggest that the maximum conversion efficiency of a central star luminosity into nebular \oiii~emission is $\sim 11\%$ \citep{1989ApJ...339...39J,1992ApJ...389...27D, 2007A&A...473..467S,2010A&A...523A..86S,2018NatAs...2..580G}. This occurs under the ideal assumption of optically thick nebula and assumes that \oiii~acts as the sole coolant. If the PNe is optically thin, then the efficiency of \oiii\ production is less, and the luminosity inferred for a PN's central star will be underestimated \citep{1992A&A...260..329M}. A high abundance of nitrogen, such as that typically found in Type I PNe \citep{1983IAUS..103..233P, 2005MNRAS.361..283P}, can also increase cooling, and lead to an underestimation of central star luminosity. Moreover, the assumption of lower limit extinction for some cases can also underestimate the luminosity. Therefore, we only consider our luminosity estimates as the lower limits. 

  \begin{figure}
   \centering
   \includegraphics[width=\hsize]{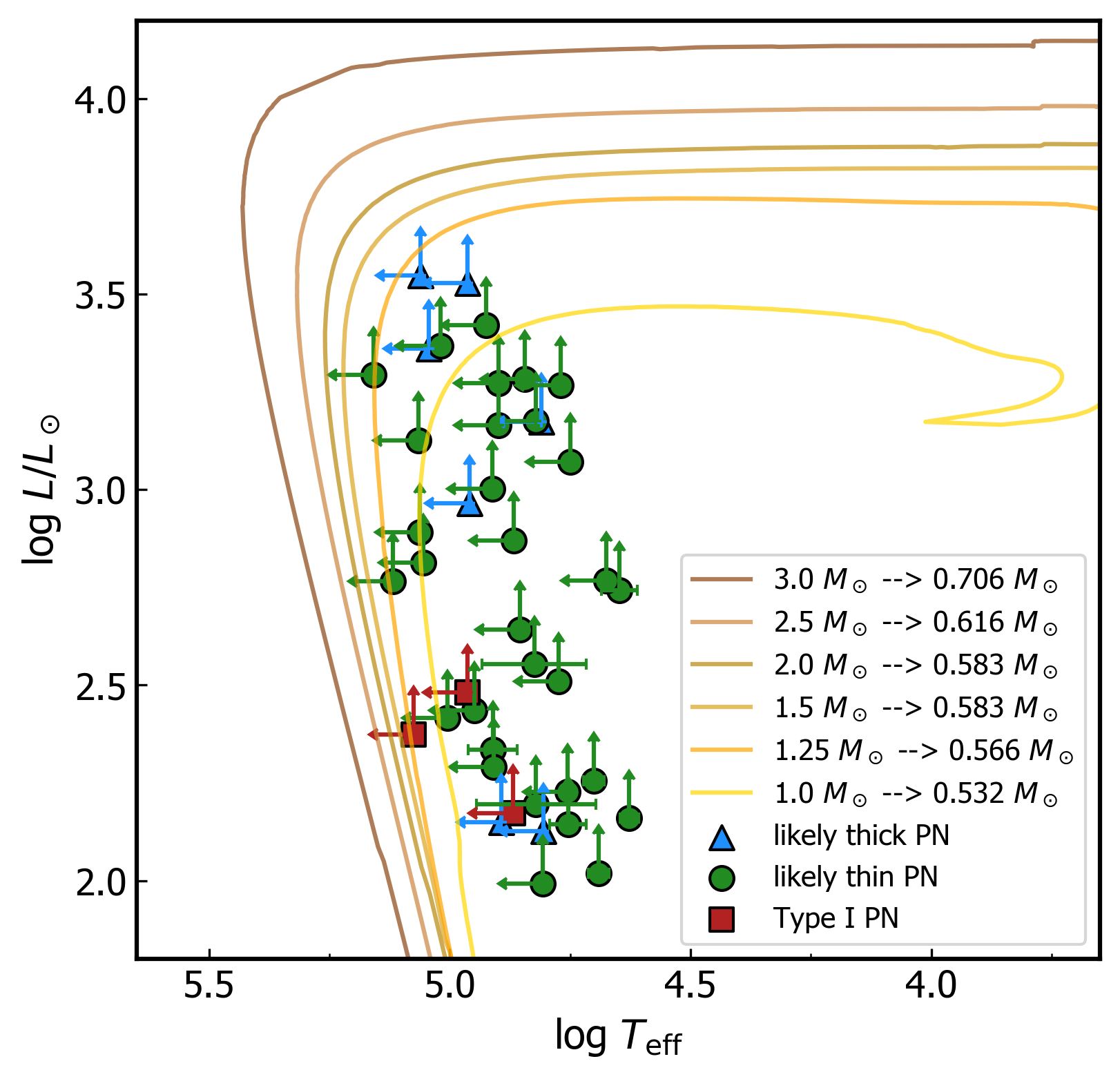}
      \caption{HR-diagram of CSPNs in NGC\,300. The evolutionary tracks are H-rich post-AGB models by \protect\citet{2016A&A...588A..25M}. The luminosity are lower limits, assuming maximum \oiii~conversion efficiency of $11\%$. For measurements within error of $\mathrm{log} \: T_{\mathrm{eff}} > 4.98$, lower limit effective temperatures are assumed. The ratio \nii/\ha~is the indicator of optical thickness, with the value less than 0.3 for \textit{more likely} optically thin. Type I PNe are classified with \nii/\ha~$> 1.0$.}
         \label{fig:central_stars}
   \end{figure}

To estimate the central stars' effective temperatures, we employed the excitation class method based on the PNe in the LMC \citep{1990ApJ...357..140D, 2010MNRAS.405.1349R}. For optically thick PNe, the excitation class temperatures are found to have an empirical correlation with temperature as derived from photo-ionisation modelling \citep{1991ApJ...367..115D, 1992ApJ...389...27D, 2010PASA...27..187R}. To employ this method, we also assume that the metallicity difference between the LMC and NGC\,300 is negligible \citep{2009ApJ...700..309B}. The revised excitation classes by \citet{2010PASA...27..187R} are defined as 
\begin{equation}
    E_{\mathrm{low}} = 0.45 \bigg{[}\frac{F(\lambda5007)}{F(\mathrm{H}\beta)}\bigg{]}
\end{equation}

\begin{equation}
    E_{\mathrm{high}} = 5.54 \bigg[\frac{F(\lambda4686)}{F(\mathrm{H}\beta)} \: + \: \mathrm{log}_{10}\frac{F(\lambda4959) + F(\lambda5007)}{F(\mathrm{H}\beta)}\bigg{]}
\end{equation}
with $E_{\mathrm{low}}$ employed for low excitation PNe ($0 < E < 5$) and $E_{\mathrm{high}}$ for medium- to high excitation PNe ($5 \leq E < 12$). The empirical relation between the excitation class and effective temperature for optically thick PNe is then defined by \citet{2010PASA...27..187R} as
\begin{equation}
    \mathrm{log} \: T_{\mathrm{eff}} = 4.439 + [0.1174(\mathrm{E})] - [0.00172(\mathrm{E}^2)]
\end{equation}
Since only the extended mode used in the MUSE-GTO dataset has the wavelength coverage to include the \heii ~line, 
for uniformity, we determine the excitation class using just the \oiii~line and \hb~line. This implies that the effective temperatures for medium- and high excitation PNe with $E \geq 5$ (or $\mathrm{log} \: T_{\mathrm{eff}} \geq 4.98$) are only lower limits. This includes the measurements, which have uncertainties beyond the condition for low-excitation PNe. Based on 6 PNe in our sample that have \heii~detection, we estimate that the effective temperatures can be underestimated by $1.5 - 3$ times if we only rely on the $E_{\mathrm{low}}$. On the other hand, if the nebula is optically thin, based on the study in the LMC, the excitation class temperatures can be overestimated by at least $50\%$ compared to the Zanstra temperatures \citep{2007ApJ...656..831V,2010PASA...27..187R}.




Both luminosity and effective temperature estimates rely on the optical thickness of the PNe. To obtain this, we adopt the criterion of \nii/\ha~$\leq 0.3$ as the condition for optically thin PNe \citep{1989ApJ...345..871K, 1989AJ.....98.1662J, 2010PASA...27..187R}. 
Since this criterion is not based on nebular modelling in our sample, we use the indication as a \textit{more likely} condition rather than a definite indicator to explain the possible limitation in our estimations. The estimated stellar parameters are presented in Figure \ref{fig:central_stars}. We include the post-AGB tracks from \citet{2016A&A...588A..25M} with a stellar metallicity of $Z_\odot = 0.01$, which is the closest to the observed value at the central area of NGC\,300 with $Z_\odot = 0.007$ \citep{2008ApJ...681..269K, 2010ApJ...712..858G}. 

A stellar population study by \citet{2020A&A...640L..19J} using the Hubble Space Telescope found young stars of $\sim 300$ Myr, AGB stars with an age between $1 - 3$ Gyr and significant number of RGB stars older than $3$ Gyr. From a single stellar evolution perspective, the stellar population of NGC\,300 can produce a PN central star mass of $\sim 0.7 \, M_\odot$ from a progenitor mass of $3.0 \, M_\odot$, which would have a main sequence lifetime of $\tau_{MS} > 320$ Myr \citep{2016A&A...588A..25M}. This implies that, theoretically, central stars within any of the stellar tracks in Figure \ref{fig:central_stars} can be expected. Unfortunately, since most of our luminosities and effective temperatures are lower limits, we are unable to put more constraints on the central star masses at this point.



Within our sample, we also identified several objects as Type I PNe \citep{1983IAUS..103..233P, 2005MNRAS.361..283P}, highly enriched in nitrogen, and classified using \nii/\ha~$>1$. These objects likely to arise from younger and more massive stars. For a progenitor mass above $\sim 2.5 \, M_\odot$, the convective envelope in the thermal pulsing AGB phase is likely to extend to the hydrogen-shell burning layer and produce ``hot bottom burning'' (HBB). This can dredge up the products of the CNO cycle to the surface, to be later expelled by the stellar wind, therefore increasing the nitrogen-to-oxygen ratio in the nebula \citep{2018MNRAS.473..241H}. Observations of PNe in M31 by \citet{2018ApJ...853...50F} put a lower limit of $\sim 2.0 \, M_\odot$ for HBB. A more thorough analysis, performed for a Type I PN in the M31 young open cluster B477-D075, yields a HBB lower mass limit of $\sim 3.4 \, M_\odot$ \citep{2019ApJ...884..115D}. This suggests that our approximation for the central star luminosities of Type I PNe is greatly underestimated. For this particular case, we argue that the assumption of \oiii~as the only coolant is not true. Since the nitrogen-to-oxygen ratio is high, the nitrogen contribution as additional coolant cannot be neglected \citep{1989ApJ...339...39J}, causing the underestimation. This might also explain why we did not see the Type I PNe at our PNLF cut-off, although they are expected to have more massive cores than the typical PNe.

More implications regarding the underlying stellar population can also be inferred from the faint end of the PNLF \citep{2010PASA...27..149C}. However, the current observational study is still limited to a relatively small sample. Recently, based on a very deep survey in M31, \citet{2021A&A...647A.130B} found that the steep rise in the number of PN fainter $M^* + 5$ mag is caused by the increased mass fraction of a population older than $5$ Gyr.  For NGC\,300, this implies that the photometry should be complete for $m_{5007} > 27$. Since our PNLF completeness breaks after $m_{5007} = 27.5$, we are unable to provide any insights on this matter at the moment. 

\subsection{Insights on the most luminous PNe}
\label{sec:cutoff}

Numerous simulation studies have been conducted to investigate the nature of the PNe at the cut-off of the luminosity function \citep{1989ApJ...339...39J,2007A&A...473..467S,2010A&A...523A..86S,2008ApJ...681..325M, 2018NatAs...2..580G, 2019AAS...23315001V}. We review some of them and compare it to our estimated properties to investigate the nature of the most luminous PNe in NGC\,300. Using the most recent post-AGB models by \citet{2016A&A...588A..25M}, simulations of the \oiii~fluxes for different progenitor mass have been performed by \citet{2018NatAs...2..580G}. They found that progenitors with the mass range between $1.5 - 3.0 \, M_\odot$ are able reach the cut-off absolute magnitude $M^* = -4.5$, assuming that the fluxes at the stellar evolution stages are maximised -- also known as maximum nebula hypothesis. It is important to note that the timescale of the $3.0 \, M_\odot$ track is too short and less likely to be observed. Additionally, they also performed a simulation with an intermediate nebula hypothesis, where the PNe are predominately opaque; this model suggests that the brightest PNe in the luminosity function will have the luminosity $\mathrm{log} \: L/L_\odot = 3.75 \pm 0.13$. For comparison, the intrinsically most luminous PN in our sample, PN H9-1, has a lower limit luminosity of $\mathrm{log} \: L/L_\odot > 3.53$. It is also indicated as more likely optically thick, which is in agreement with the simulation. Again, we note that this assumes the ideal $11\%$ maximum efficiency. For example, based on the chemical abundance analysis, the bright PNe in M31 exhibit less conversion efficiency \citep{1999ApJ...515..169J, 2012ApJ...753...12K}. Therefore, the actual central star luminosity is likely to be brighter. 

Simulation of \oiii~flux evolution has also been conducted by \citet{2007A&A...473..467S}, who employed 1-dimensional radiative-hydrodynamical simulations for the nebulae. They calculated that the most luminous PNe that populate the PNLF cut-off will achieve their maximum luminosity at $\mathrm{log} \, T_{\mathrm{eff}} = 5.00$ K and spend $\sim 500$ years in this phase. For PN H9-1, the lower limit temperature is $\mathrm{log} \, T_{\mathrm{eff}} > 4.97$. They also suggest that UV- to \oiii~flux conversion process happens most efficiently for central star mass of $\sim 0.62 \, M_\odot$, if the nebular shell remains optically thick during the evolution. Referring to the post-AGB models by \citet{2016A&A...588A..25M}, the initial mass of the progenitor star would be $\sim 2.5 \, M_\odot$, and the object would spend less than $1000$ years before entering the white dwarf cooling sequence. 

Similarly, hydrodynamical models have been used to investigate PNe in nearby galaxies by \citet{2010A&A...523A..86S}. In these simulations, it was found that central star masses greater than $0.65 \, M_\odot$ do not exist at the PNLF cut-off. This also supports the result from \citet{2018NatAs...2..580G}, in that a progenitor mass of $3.0 \, M_\odot$ for PNe is not expected. This also agrees with the progenitor masses between $2.0 - 2.5 \, M_\odot$ for the bright PNe in NGC\,300 predicted by \citet{2013A&A...552A..12S}, despite the concerns we mentioned regarding their spectroscopic fluxes in Section \ref{section-photometry}. They derived the progenitor masses using the stellar tracks of \citet{1995A&A...299..755B}, which evolve slower than the recent models of \citet{2016A&A...588A..25M}. This implies the possibility of less massive progenitors if the new stellar tracks are adopted, which is however beyond the scope of our current study.


Recently, the properties of luminous PNe near the PNLF cut-off of M31 have been studied by \citet{2018ApJ...863..189D} for the bulge, and by \citet{2022A&A...657A..71G} for the disk. For the disk, it was found that the four brightest PNe have an average progenitor mass of $1.5 \, M_\odot$, which is lower than the values predicted by \citet{2007A&A...473..467S}, but still in agreement with \citet{2018NatAs...2..580G}. \citet{2022A&A...657A..71G} also measure a relatively low average extinction of the PNe with $c(\mathrm{H}\beta) \sim 0.1$. This means that the PNe originated from an older stellar population, although the disk of M31 also exhibits star forming regions. 

In contrast, in the older population of the bulge of M31, \citet{2018ApJ...863..189D} found that the brightest PN have a central star mass $> 0.66 \, M_\odot$, which means progenitor masses of $> 2.5 \, M_\odot$. This is found for cases with high extinction, one even reaching $c(\mathrm{H}\beta) \sim 0.6$, with the average of $c(\mathrm{H}\beta) \sim 0.3$ for 23 PNe. Current simulations do not predict such massive central stars to be observable, if they exist at all \citep{2010A&A...523A..86S,2018NatAs...2..580G}. In old systems, the most luminous PNe are suggested to be products from of blue stragglers -- stars that result from a merger during the main sequence \citep{2005ApJ...629..499C}, or symbiotic nebula \citep{2006ApJ...640..966S}. However, both scenarios still do not predict such massive central stars to exist. While the bright \ha~background might overestimate the measured extinction, which can lead to the overestimation of luminosity and subsequently the progenitor mass, \citet{2018ApJ...863..189D} in fact did their measurement with an IFU. Their sky subtraction was based on a PSF model, which was claimed to be accurate within $10\%$. 

Lately, \citet{2021PASP..133i3002U} suggested that the extinction measurement should be solved iteratively, considering the dependency of \ha/\hb~ratio on the electron temperature ($T_e$) and electron density ($n_e$). Assuming those two parameter as constants would increase the uncertainty of the extinction, and subsequently the stellar parameters. They demonstrate this approach by reanalysing the M31 disk PNe, worked by \citet {2022A&A...657A..71G}, and found that the iterative approach yields an average progenitor mass of $2.2 \, M_\odot$, instead of $1.5 \, M_\odot$ for the four brightest PNe \citep{2022arXiv221007091U}. While the extinction does not necessarily affect the $T_e$ and $n_e$, it may compromise the ionic and elemental abundance analysis \citep{2021PASP..133i3002U,2022arXiv221007091U}. Since we also assume constant $T_e$ and $n_e$ for our parameters, we are not excluded from this problem. However, as we are missing the diagnostic lines in the blue spectral region and the ones within MUSE wavelength coverage are below the detection limit, we are unable to put constraints on the $T_e$ and $n_e$.

It would be interesting to repeat the exercise of modelling PN spectra on the basis of improved IFU observations that we believe are superior to slit-based spectroscopy in controlling systematic errors, with the more recent stellar evolution tracks and more careful plasma diagnostics. The future BlueMUSE instrument for the VLT \citep{2019arXiv190601657R} will offer the capability with a wavelength coverage down to the atmospheric limit in the UV, which includes the necessary nebular lines for such study.

\section{Conclusions}

We analyse 44 fields, obtained with the MUSE instrument to find PNe and construct the PNLF. Using the differential emission line filter \citep[DELF,][]{2021arXiv210501982R}, we identified more than 500 point sources in \oiii, 107 of which were designated as PNe based on spectral classification with the aid of the BPT-diagram \citep{1981PASP...93....5B}. The \oiii~magnitudes for the PNe were obtained using DAOPHOT aperture photometry \citep{1987PASP...99..191S} with aperture corrections. With the sample completeness at $m_{5007} = 27$ for most fields, we constructed the PNLF, which exhibits the dip that has been observed in other star forming galaxies \citep{2002AJ....123..269J, 2010PASA...27..149C, 2010MNRAS.405.1349R}. To derive the distance, we employed the maximum likelihood estimation method \citep{1989ApJ...339...53C} to yield a most likely distance modulus $(m-M)_0 = 26.48^{+0.11}_{-0.26}$ ($d = 1.98^{+0.10}_{-0.23}$ Mpc). For PNe, that are isolated from surrounding emission line sources, and that exhibit bright enough Balmer lines, we measured their extinction. We estimated parameters of the central stars using the extinction corrected fluxes in an attempt to track their origin from the underlying stellar population. We discuss the accuracy of our distance measurement, the effect of local dust for our PNe extinction measurements, and the properties of the most luminous PNe in our sample. The conclusions are as follows:

   \begin{enumerate}
      
      \item The PNLF distance measurement to NGC\,300 is improved with our method and is in excellent agreement with Cepheids and TRGB distances. This is due to the spectral information and spatial resolution of MUSE, that provides a higher PN detection per area, better classification, and accurate photometry.
      
      \item With a limited sample, distance determination based on the minimisation technique is very dependent on the binning. Although coarse binning might provide a better apparent shape of the luminosity function for fitting, it can introduce an unintended systematic shift. Moreover, the details of the PNLF shape, which can provide insights on the stellar population, are also smeared out. 
      
      \item The extinction derived for the PNe cannot be disentangled completely from the local dust lane extinction within the galaxy. However, with the spatial resolution of MUSE, we were able to resolve several PNe that are likely obstructed by dust lanes. Any attempt to link the internal extinction and the underlying stellar population requires a quantitative technique to separate the local and internal PNe extinction. 
      
      
      \item We found a few Type I PNe, that evolved from main sequence mass $>2.5 \, M_\odot$. Their luminosities are likely underestimated due to the high abundance of nitrogen that serves as a competing coolant with oxygen. They do not populate our PNLF cut-off.  
      
   \end{enumerate}
   
   
   With these results, and other works reported in the literature, we feel encouraged to further develop the IFU observing technique with MUSE to study extragalactic PNe. One of the inherent parameters that we have as yet not utilised is the radial velocity of individual PNe that comes for free as a by-product of the analysis. It will be interesting to find out whether the kinematics can provide hints as to the membership in different populations in NGC\,300. Such study was recently done for other disc galaxies: NGC\,628 \citep{2018MNRAS.476.1909A}, NGC\,6946 \citep{2021MNRAS.500.3579A}, and M31 \citep{2019A&A...624A.132B}.  In the interest of understanding the physical parameters of the PNe, we are currently dependant on the ideal assumption of \oiii~maximum conversion and excitation classes to derive the central star parameters, which is not ideal, especially when most cases have no \heii~coverage. Better constraints on the luminosities and effective temperatures are obtainable through nebular abundance modelling. However, our current wavelength coverage of the MUSE instrument limit us to explore this possibility. Future IFUs, that are optimised in the blue wavelength, such as BlueMUSE \citep{2019arXiv190601657R}, will play an important role and allow us to gain more understanding about PNe in the nearby galaxies beyond the Local Group, getting us closer to comprehend the underlying physics behind the constancy of PNLF cut-off across galaxies.

\begin{acknowledgements}
      We thank the anonymous referee for a critical reading of the manuscript and helpful suggestions to improve the quality of this paper. Part of this work was supported by the German BMBF program Unternehmen Region, grant 03Z22AN11. PMW gratefully acknowledges support by the BMBF from the ErUM program (project VLT-BlueMUSE, grant 05A20BAB). N. Castro gratefully acknowledges funding from the Deutsche Forschungsgemeinschaft (DFG) – CA 2551/1-1.
\end{acknowledgements}

\bibliographystyle{aa}
\bibliography{pnlf_ngc300.bbl}

\begin{appendix}

\section{Aperture correction}
\label{appendix:apcor}

The radial profile of a PSF is best modelled with a Moffat function, as a Gaussian often does not accurately match the wings of the PSF \citep{2002AJ....124..266P, 2013A&A...549A..71K}. Moreover, flux measurements using a discrete aperture are not able to collect all of the flux from the PSF wings. To recover the lost flux and obtain accurate photometry, we therefore need to apply an aperture correction to our measurements. In order to do this, we need at least one star in a given field as a reference for the observation's PSF\null. We examined 3-4 objects to infer the average PSF FWHM of the frame and chose the best star for the aperture correction. Moreover, we also examined the behaviour of the PSF across wavelengths, as the PSF is expected to be more extended in the blue, and to show a monotonic decrease of the FWHM toward the red \citep{1966JOSA...56.1372F, 1978JAtS...35.2236B, 2013A&A...549A..71K}. 

To obtain the aperture correction value, we collected the flux of the reference star using a large aperture radius of 2\farcs4 (or 12 spaxels), assuming that almost all of the flux will be recorded \citep{1989PASP..101..616H}. Then, by taking the flux of the same star with the aperture size employed for the PNe, we were able to obtain the correction value by taking the ratio of the two fluxes. We then applied this constant to all PNe measurements within the field. 

For the Balmer lines, we have to make sure that both lines are corrected in a consistent manner, especially with respect to the wavelength dependence of the PSF\null. Since the seeing at the telescope is decreasing monotonically with wavelength, the FWHM for \hb~is expected to be larger than the one for \ha, thus changing the aperture correction. The reference stars in each field therefore have to be well behaved across this wavelength range which was found to not always be the case. We found several apparent point sources that unexpectedly exhibit an increasing PSF FWHM trend to the red. Closer inspection revealed that stellar crowding with luminous red stars, e.g., M giants and carbon stars, were responsible for this problem. For Balmer line corrections, we decided to discard the problematic stars as useful PSF references.


  \begin{figure}
   \centering
   \includegraphics[width=\hsize]{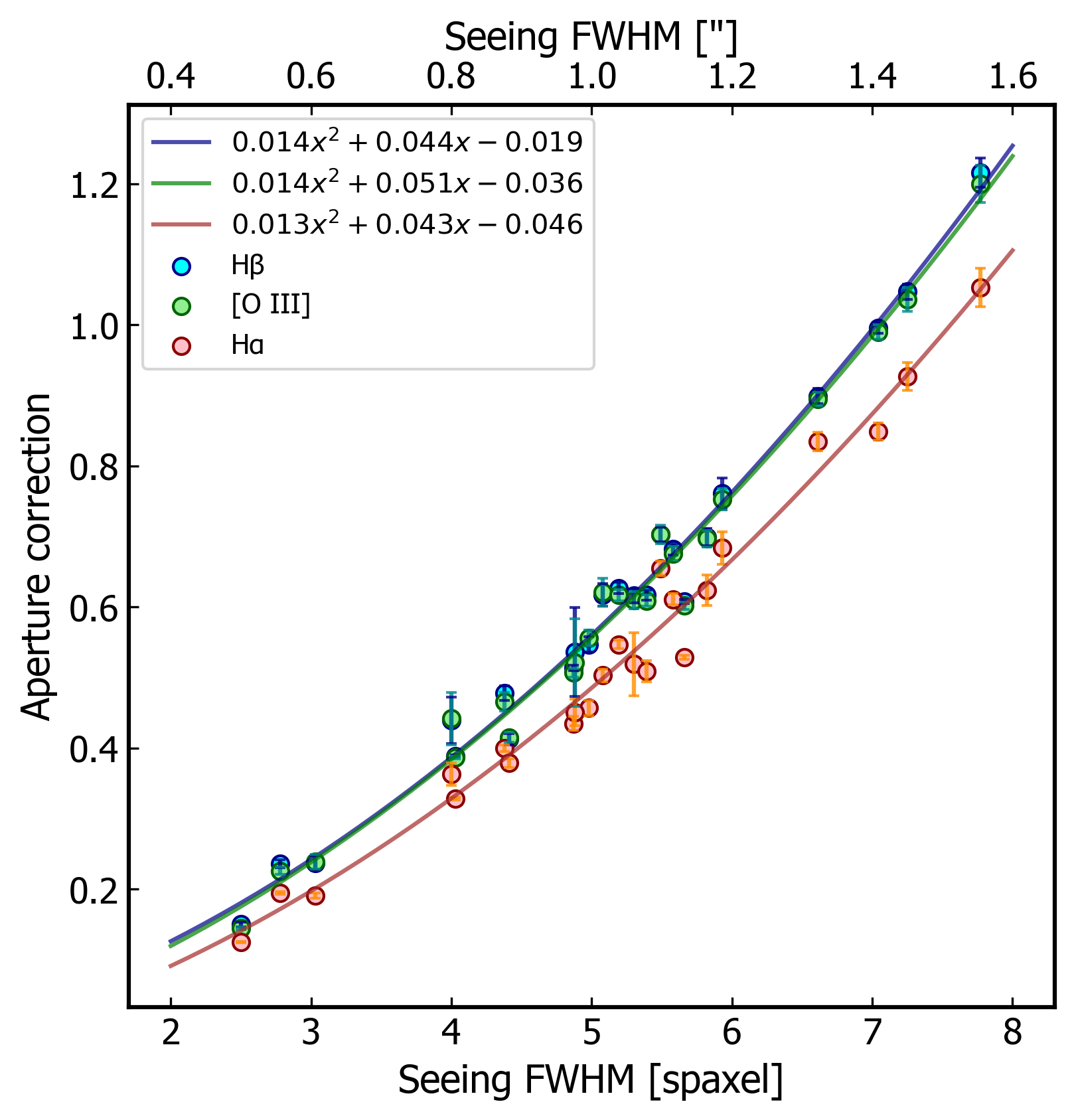}
      \caption{Empirical polynomial relation between the seeing FWHM and aperture correction for \hb, \oiii, and \ha. The relation is derived using 23 stars from different fields.}
         \label{fig:aperture_correction}
   \end{figure}

As an alternative, we used the brightest PNe that happen to be sufficiently isolated from nearby diffuse gas and \hii~regions. We then used the PNe's image profile at the wavelengths of the strong lines of \oiii~and \ha, while assuming a negligible difference between the PSF at 5007\,Å and \hb. Unfortunately, we found that some of our fields have neither a well behaved star, nor bright isolated PNe, so another alternative was needed. Using the best reference stars from different fields and seeing conditions, we derived a simple polynomial relation between the seeing FWHM and the aperture correction value for \hb, \oiii, and \ha; these curves are presented in Figure \ref{fig:aperture_correction}. 

We also confirmed that the difference between PSF at \hb~and \oiii~is not significant, as the polynomial fit is almost identical for the two wavelengths.  However, it is important to mention that the relation is only derived using a limited sample of 23 stars. It is not possible to determine the true distribution of this relation and identify the variables that affect it. While this is worthwhile for further investigation, we will not explore it in the current study. We employed the empirical relation as the final alternative, after the bright PN method and the main reference star method. From the 50 PNe that are within the \ha~threshold, we corrected 16 PNe with the reference star method, 18 PNe with the bright PN method, and 16 PNe with the empirical relation.

In future studies, the aperture correction can be improved. Firstly, the uncertainties can be minimised under excellent seeing conditions, ideally 0\farcs6 PSF FWHM at the wavelength of \oiii, that can be achieved using the adaptive optics mode of MUSE. In cases where no field star is available to serve as a PSF reference, modelling the wavelength and seeing dependant PSF on the basis of instrumental data from the adaptive optics control software may provide a way out \citep{Fusco+2020}.

\section{Extinction uncertainties}
\label{extinction_uncertainty}

The main uncertainty of our extinction measurement is the aperture correction. Since we have only a limited number of objects observed with each correction method, every PN has its own uncertainty, making it difficult to we derive a proper statistical error. As an alternative, we estimate the error based on the comparison of extinction calculated from different aperture correction methods. To perform this, we considered PN candidates that were measured with a well behaved reference star in the field, and a bright PN in the same field. The comparison is presented in Table \ref{apcor_compare}. 

\begin{table}
\caption{\label{apcor_compare}Comparison for extinction values in \oiii~derived using different aperture correction methods. The preferred extinction values are marked in bold.}
\centering
\begin{tabular}{lcccccc}
\hline\hline
ID & $A_0$ & $A_1$ & $A_2$ & $A_3$ & $\Delta A$\\
\hline \\[-3.2mm]
E-2 & 0.542 & 0.589\tablefootmark{a} & \textbf{0.353} & 0.457 & 0.189 \\[1mm]
E-11 & 0.615 & 0.663\tablefootmark{a} & \textbf{0.445} & 0.531 & 0.170 \\[1mm]
P-2 & 1.342 & \textbf{1.229} & 1.093\tablefootmark{b} & 1.252 & 0.113 \\[1mm]
H9-1 & 0.358 & \textbf{0.147} & 0.257 & 0.223 & 0.211 \\[1mm]
H9-6 & 0.490 & \textbf{0.279} & -0.027\tablefootmark{b} & 0.354 & 0.211 \\[1mm]
H10-2 & 0.694 & \textbf{0.420} & 0.344 & 0.506 & 0.274 \\[1mm]
H11-1 & 0.570 & \textbf{0.494} & 0.412 & 0.413 & 0.076 \\[1mm]
H11-2 & 1.440 & \textbf{1.364} & 1.276 & 1.283 & 0.076 \\[1mm]
L9-8 & 0.633 & \textbf{0.466} & 0.463 & 0.410 & 0.167 \\
\hline
\end{tabular}
\tablefoot{$A_0$ -- no aperture correction; $A_1$ -- reference star method; $A_2$ -- bright PN method; $A_3$ -- empirical relation method.\\
\tablefoottext{a}{bad reference star}\\
\tablefoottext{b}{uniform diffuse \ha~background}\\[-5mm]
}
\end{table}

Since we expect the aperture correction at \hb~to be larger than the one for \ha, the application of this factor will reduce the inferred extinction, as \hb~appears in the denominator of equation~(\ref{eq:extinction}).  In field E, where the initially selected reference star shows the unusual trend of an increasing PSF width to the red, we computed the extinction to be {\em larger} after the aperture correction, prompting us to restrict the reference star method only to cases where a well behaved star is available in the field. Moreover, we also see that the use of PNe that are not completely isolated from the ambient gas tends to underestimate the extinction, if compared to other aperture correction methods. Based on our choice of priority of the methods, marked as bold in Table \ref{apcor_compare}, the difference between the extinction with and without aperture correction ($\Delta A$) is always larger than the difference between extinction values derived using various aperture correction methods. Therefore, we decided to select the $\Delta A$ as our error estimates. In cases where the error estimates exceed nonphysical negative extinction, the lower limit of the uncertainty is assumed until zero extinction.

We should note, as discussed in Section \ref{sec:cutoff}, that assuming a constant electron temperature {$T_e$} and {$n_e$} also introduce uncertainties \citep{2021PASP..133i3002U, 2022arXiv221007091U}. Since we did not have the capability to measure {$T_e$} and {$n_e$}, we did not include this aspect in our measurement error. 

\section{MUSE observation fields}
\label{appendix:fields}

The details of the MUSE fields, both the MUSE-GTO and ML20 \citep{2020ApJ...891...25M, 2021MNRAS.508.5425M}, can be referred in Table \ref{tab:observation-museGTO}. We also include the seeing in \oiii, which obtained based on the average FWHM of 3-4 point sources, preferably stars, in each field.

\begin{table*}[h!]
\caption{\label{tab:observation-museGTO} The MUSE observation fields for this work. The upper part represent the MUSE-GTO data and the lower part the ML20 data.}
\centering

\begin{tabular}{ccccc}
\hline\hline
Field & RA(2000) & DEC(2000) & Observation date & FWHM$_{5007}$ ["] \\
\hline \\[-3.2mm]
A & 00:54:53.62 & -37:41:05.1 & 2018-10-15 & 0\farcs67 \\ [1mm]
B & 00:54:48.54 & -37:41:05.3 & 2018-10-15 & 0\farcs69 \\ [1mm]
C & 00:54:43.49 & -37:41:05.1 & 2018-11-13 & 0\farcs82 \\ [1mm]
D & 00:54:42.32 & -37:42:05.1 & 2015-08-24 & 0\farcs79 \\ [1mm]
E & 00:54:48.17 & -37:42:13.7 & 2015-09-13 & 0\farcs60 \\ [1mm]
I & 00:54:37.08 & -37:40:52.6 & 2014-10-30 & 0\farcs66 \\ [1mm]
J & 00:54:39.49 & -37:39:50.4 & 2014-11-26 & 0\farcs82 \\ [1mm]
P & 00:54:24.00 & -37:36:29.0 & 2016-09-03 & 0\farcs59 \\ [1mm]
Q & 00:54:22.00 & -37:37:47.0 & 2016-09-03 & 0\farcs55 \\ [1mm]
\hline \\[-3.2mm]
H1 & 00:54:59.83 & –37:39:42.0 & 2016-10-01 & 1\farcs00 \\[1mm]
H2 & 00:54:55.40 & -37:39:17.0 & 2016-10-01 & 1\farcs10 \\[1mm]
H3 & 00:54:50.99 & –37:38:51.8 & 2016-10-01 & 0\farcs87 \\[1mm]
H4 & 00:54:46.55 & –37:38:26.7 & 2016-10-04 & 1\farcs19 \\[1mm]
H5 & 00:55:06.51 & –37:41:25.5 & 2016-10-05 & 1\farcs49 \\[1mm]
H6 & 00:55:02.08 & –37:41:00.9 & 2016-10-05 & 1\farcs18 \\[1mm]
H7 & 00:54:57.65 & –37:40:35.7 & 2016-10-05 & 1\farcs30 \\[1mm]
H8 & 00:54:53.22 & –37:40:10.3 & 2016-10-05 & 1\farcs06 \\[1mm]
H9 & 00:54:48.81 & –37:39:45.4 & 2016-11-07 & 0\farcs82 \\[1mm]
H10 & 00:54:44.37 & –37:39:20.3 & 2016-11-08 & 0\farcs96 \\[1mm]
H11 & 00:54:39.95 & –37:38:55.1 & 2016-11-08 & 0\farcs86 \\[1mm]
H12 & 00:55:04.35 & –37:42:19.4 & 2016-11-08 & 0\farcs88 \\[1mm]
H13 & 00:54:59.90 & –37:41:54.7 & 2016-11-08 & 0\farcs94 \\[1mm]
H14 & 00:54:55.47 & –37:41:29.3 & 2016-11-08 & 1\farcs14 \\[1mm]
H15 & 00:54:57.70 & –37:42:48.2 & 2016-11-08 & 1\farcs23 \\[1mm]
H16 & 00:54:55.52 & –37:43:42.0 & 2016-12-19 & 1\farcs02 \\[1mm]
H17 & 00:54:51.08 & –37:43:16.8 & 2016-12-23 & 1\farcs02 \\[1mm]
L1 & 00:55:08.69 & –37:40:32.3 & 2016-12-23 & 1\farcs13 \\[1mm]
L2 & 00:55:04.26 & –37:40:06.8 & 2016-12-23 & 1\farcs01 \\[1mm]
L3 & 00:54:42.13 & –37:38:01.3 & 2016-12-24 & 1\farcs08 \\[1mm]
L4 & 00:54:51.04 & –37:41:04.2 & 2016-12-26 & 1\farcs11 \\[1mm]
L5 & 00:54:46.63 & –37:40:38.9 & 2017-01-02 & 1\farcs42 \\[1mm]
L6 & 00:54:42.19 & –37:40:13.7 & 2017-01-02 & 1\farcs50 \\[1mm]
L7 & 00:54:37.77 & –37:39:48.5 & 2017-01-04 & 1\farcs49 \\[1mm]
L8 & 00:55:02.15 & –37:43:13.1 & 2018-07-03 & 0\farcs77 \\[1mm]
L9 & 00:54:53.27 & –37:42:22.6 & 2017-01-05 & 1\farcs16 \\[1mm]
L10 & 00:54:48.84 & –37:41:57.9 & 2017-01-06 & 0\farcs84 \\[1mm]
L11 & 00:54:44.43 & –37:41:32.7 & 2017-01-06 & 0\farcs84 \\[1mm]
L12 & 00:54:39.99 & –37:41:07.4 & 2017-01-06 & 0\farcs96 \\[1mm]
L13 & 00:54:35.57 & –37:40:42.2 & 2017-01-07 & 0\farcs84 \\[1mm]
L14 & 00:54:59.95 & –37:44:06.8 & 2017-01-07 & 1\farcs01 \\[1mm]
L15 & 00:54:46.64 & –37:42:51.6 & 2017-01-16 & 1\farcs00 \\[1mm]
L16 & 00:54:42.21 & –37:42:26.4 & 2017-01-27 & 1\farcs16 \\[1mm]
L17 & 00:54:37.79 & –37:42:00.9 & 2018-07-04 & 1\farcs01 \\[1mm]
L18 & 00:54:33.37 & –37:41:36.2 & 2018-07-04 & 0\farcs95 \\
\hline
\end{tabular}

\end{table*}

\section{MUSE-PN catalogue}
\label{appendix:catalogue}

The MUSE-PN catalogue of NGC\,300 is presented in Table \ref{tab:catalogue}. MUSE-GTO coordinates (accuracy of $\sim0\farcs1$) are presented if available, otherwise ML20 coordinates (accuracy of $\sim 3\arcsec$) are provided. The table will be available through the CDS Archive and will also include the columns for aperture corrected line fluxes of \hb, \oiii, \ha, \nii, \sii, \siiright, and the classification remarks.

\longtab[1]{

\begin{longtable}{c c c c c c c c c c}
\caption{MUSE-PN catalogue of NGC\,300} \label{tab:catalogue}\\
\hline
\hline
No & ID$_{\mathrm{GTO}}$ & ID$_{\mathrm{McLeod}}$ & ID$_{\mathrm{PE12}}$ & RA(2000) & DEC(2000) & $m_{5007} $ & c(H$\beta$) & $\mathrm{log} \: L$ [$L_\odot$]\tablefootmark{a} &$\mathrm{log} \: T_\mathrm{eff}$ [K]\\ 
\hline \\ [-3.2mm]
\endfirsthead
\caption{continued.} \\
\hline
No & ID$_{\mathrm{GTO}}$ & ID$_{\mathrm{McLeod}}$ & ID$_{\mathrm{PE12}}$ & RA(2000) & DEC(2000) & $m_{5007} $ & c(H$\beta$) & $\mathrm{log} \: L$ [$L_\odot$]\tablefootmark{a} &$\mathrm{log} \: T_\mathrm{eff}$ [K]\\
\hline \\ [-3.2mm]
\endhead
\hline
\endfoot
\hline 
\endlastfoot
1 & - & H9-1 & PN35 & 0:54:48.44 & -37:39:48.39 & 22.10$\pm$0.06 & 0.04$\pm$0.08 & 3.53 & 4.97\tablefootmark{*}\\ [1mm]
2 & - & H7-3 & PN53 & 0:54:56.54 & -37:40:28.99 & 22.48$\pm$0.06 & 0.00$\pm$0.07 & 3.37 & 5.02\tablefootmark{*}\\ [1mm]
3 & - & H7-2 & PN58 & 0:54:58.41 & -37:40:44.29 & 22.68$\pm$0.06 & 0.11$\pm$0.08 & 3.36 & 5.05\tablefootmark{*}\\ [1mm]
4 & C-7 & L11-2 & PN25 & 0:54:44.45 & -37:41:29.36 & 22.85$\pm$0.06 & 0.03$\pm$0.10 & 3.17 & 4.90\tablefootmark{*}\\ [1mm]
5 & E-2 & L7-8 & PN14 & 0:54:38.95 & -37:39:43.26 & 22.86$\pm$0.06 & 0.10$\pm$0.07& 3.29 & 4.74\tablefootmark{*}\\ [1mm]
6 & A-11 & H14-7 & PN51 & 0:54:55.34 & -37:41:28.34 & 22.98$\pm$0.06 & 0.10$\pm$0.05& 3.27 & 4.80\tablefootmark{*}\\ [1mm]
7 & - & L6-7 & PN22 & 0:54:42.24 & -37:40:04.82 & 22.99$\pm$0.06 & 0.20$\pm$0.09& 3.29 & 5.16\tablefootmark{*}\\ [1mm]
8 & A-23 & H14-8 & PN48 & 0:54:54.95 & -37:41:32.88 & 23.01$\pm$0.06 & 0.42$\pm$0.05& 3.54 & 5.06\tablefootmark{*}\\ [1mm]
9 & I-2 & L11-11 & PN24 & 0:54:43.75 & -37:41:51.52 & 23.18$\pm$0.06 & 0.02$\pm$0.04& 3.13 & 5.07\tablefootmark{*}\\ [1mm]
10 & - & L9-8 & PN40 & 0:54:52.16 & -37:42:43.27 & 23.20$\pm$0.06 & 0.13$\pm$0.07& 3.18 & 4.82\tablefootmark{*}\\ [1mm]
11 & P-2 & - & - & 0:54:25.37 & -37:36:29.93 & 23.25$\pm$0.06 & 0.38$\pm$0.05 & 3.42 & 4.92\tablefootmark{*}\\ [1mm]
12 & - & H6-3 & PN69 & 0:55:04.25 & -37:40:52.20 & 23.29$\pm$0.06 & 0.26$\pm$0.07& 3.27 & 4.77\tablefootmark{*}\\ [1mm]
13 & - & H2-6 & PN45 & 0:54:54.03 & -37:39:28.14 & 23.36$\pm$0.06 & 0.17$\pm$0.11& 3.08 & 4.75\tablefootmark{*}\\ [1mm]
14 & - & H1-8 & PN54 & 0:54:57.03 & -37:39:44.09 & 23.39$\pm$0.06 & 0.19$\pm$0.06& 3.17 & 4.81\tablefootmark{*}\\ [1mm]
15 & - & L2-3 & PN66 & 0:55:02.42 & -37:39:54.64 & 23.40$\pm$0.06 & 0.05$\pm$0.08& 3.00 & 4.91\tablefootmark{*}\\ [1mm]
16 & - & H10-2 & PN23 & 0:54:43.57 & -37:39:35.80 & 23.65$\pm$0.06 & 0.13$\pm$0.09& 2.96 & 4.96\tablefootmark{*}\\ [1mm]
17 & - & H7-7 & - & 0:54:58.44 & -37:41:09.84 & 23.82$\pm$0.06 & 0.00\tablefootmark{*}& 2.89 & 5.06\tablefootmark{*}\\ [1mm]
18 & - & H14-19 & PN60 & 0:54:58.34 & -37:41:16.14 & 23.88$\pm$0.06 & 0.05$\pm$0.11& 2.77 & 4.68\tablefootmark{*}\\ [1mm]
19 & - & H6-5 & PN65 & 0:55:02.00 & -37:40:28.21& 23.93$\pm$0.06 & 0.10$\pm$0.07& 2.87 & 4.86\tablefootmark{*}\\ [1mm]
20 & E-11 & L7-2 & PN12 & 0:54:37.93 & -37:40:14.29 & 24.21$\pm$0.06 & 0.12$\pm$0.07& 2.77 & 5.12\tablefootmark{*}\\ [1mm]
21 & A-31 & L4-6 & PN42 & 0:54:53.28 & -37:40:54.37 & 24.45$\pm$0.06 & 0.24$\pm$0.05& 2.81 & 5.05\tablefootmark{*}\\ [1mm]
22 & - & H14-18 & PN61 & 0:54:58.49 & -37:41:13.81 & 24.49$\pm$0.06 & 0.00$\pm$0.07& 2.55 & 4.82$\pm$0.11\\ [1mm]
23 & - & L17-6 & - & 0:54:38.46 & -37:42:29.27 & 24.53$\pm$0.06 & - & - & -\\ [1mm]
24 & Q-1 & - & - & 0:54:22.61 & -37:37:52.67 & 24.64$\pm$0.06& 0.11$\pm$0.03 & 2.64 & 4.85\tablefootmark{*}\\ [1mm]
25 & I-5 & L15-2 & - & 0:54:44.81 & -37:42:27.13 & 24.96$\pm$0.06 & 0.04$\pm$0.04& 2.43 & 4.95\tablefootmark{*}\\ [1mm]
26 & - & H12-14 & - & 0:55:04.02 & -37:41:51.29 & 25.01$\pm$0.06 & 0.00\tablefootmark{*}& 2.41 & 5.00\tablefootmark{*}\\ [1mm]
27 & C-6 & L6-6 & PN21 & 0:54:41.95 & -37:40:43.32 & 25.07$\pm$0.06 & 0.29$\pm$0.10& 2.51 & 4.77\tablefootmark{*}\\ [1mm]
28 & - & H11-2 & PN10 & 0:54:37.65 & -37:39:03.73 & 25.07$\pm$0.06 & 0.42$\pm$0.03& 2.74 & 4.65$\pm$0.04\\ [1mm]
29 & B-1 & L4-8 & PN37 & 0:54:49.46 & -37:40:42.51 & 25.13$\pm$0.06& 0.13$\pm$0.04& 2.45 & 4.96\tablefootmark{*}\\ [1mm]
30 & - & H12-13 & - & 0:55:04.02 & -37:41:52.26 & 25.21$\pm$0.06 & 0.00\tablefootmark{*}& 2.33& 4.86\tablefootmark{*}\\ [1mm]
31 & - & H1-3 & - & 0:55:00.90 & -37:39:39.74 & 25.23$\pm$0.06 & 0.00\tablefootmark{*} & 2.33 & 5.08\tablefootmark{*}\\ [1mm]
32 & - & L5-8 & PN27 & 0:54:45.05 & -37:40:28.67 & 25.32$\pm$0.06 & 0.00\tablefootmark{*}& 2.29 & 4.91\tablefootmark{*}\\ [1mm]
33 & - & H6-9 & PN62 & 0:54:59.83 & -37:41:00.24 & 25.41$\pm$0.06 & 0.00\tablefootmark{*} & 2.25 & 4.70$\pm$0.02\\ [1mm]
34 & - & H9-6 & PN29 & 0:54:45.85 & -37:39:58.04 & 25.52$\pm$0.06 & 0.08$\pm$0.08& 2.20 & 4.70\tablefootmark{*}\\ [1mm]
35 & - & H6-2 & PN72 & 0:55:05.13 & -37:40:45.33 & 25.57$\pm$0.06 & - & - & -\\ [1mm]
36 & A-25 & H14-4 & - & 0:54:55.04 & -37:41:17.31 & 25.59$\pm$0.06 & - & - & - \\ [1mm]
37 & C-2 & L11-1 & - & 0:54:45.79 & -37:41:30.80 & 25.60$\pm$0.06 & 0.21$\pm$0.10& 2.23 & 4.76\tablefootmark{*}\\ [1mm]
38 & - & H9-11 & PN29 & 0:54:49.31 & -37:40:19.15 & 25.65$\pm$0.06 & 0.00\tablefootmark{*}& 2.16 & 4.63$\pm$0.02\\ [1mm]
39 & - & H16-1 & PN52 & 0:54:56.02 & -37:43:14.52 & 25.66$\pm$0.07 & 0.00\tablefootmark{*}& 2.15 & 4.89\tablefootmark{*}\\ [1mm]
40 & - & H12-2 & PN67 & 0:55:03.18 & -37:42:08.97 & 25.69$\pm$0.06 & 0.00\tablefootmark{*}& 2.14 & 4.75$\pm$0.04\\ [1mm]
41 & - & L14-2 & - & 0:55:01.17 & -37:44:36.29 & 25.72$\pm$0.07 & 0.00\tablefootmark{*}& 2.13 & 4.81\tablefootmark{*}\\ [1mm]
42 & - & H9-2 & PN31 & 0:54:48.11 & -37:39:41.62 & 25.78$\pm$0.06 & 0.00\tablefootmark{*}& 2.11 & 4.86\tablefootmark{*}\\ [1mm]
43 & - & L9-2 & PN43 & 0:54:53.45 & -37:41:56.10 & 25.80$\pm$0.07 & - & - & -\\ [1mm]
44 & - & H9-4 & PN38 & 0:54:49.77 & -37:39:48.14 & 25.84$\pm$0.06 & - & - & - \\ [1mm]
45 & E-4 & L7-7 & PN13 & 0:54:38.17 & -37:39:41.42 & 25.92$\pm$0.06 & 0.02$\pm$0.07& 2.00 & 4.81\tablefootmark{*}\\ [1mm]
46 & - & H8-8 & - & 0:54:51.02 & -37:40:03.35 & 25.97$\pm$0.07& - & - & - \\ [1mm]
47 & - & H6-1 & PN68 & 0:55:04.00 & -37:40:41.44 & 26.00$\pm$0.06 & 0.00\tablefootmark{*} & 2.02 & 4.69$\pm$0.02\\ [1mm]
48 & C-5 & L12-1 & - & 0:54:42.12 & -37:41:00.04 & 26.04$\pm$0.06 & 0.00\tablefootmark{*} & - & -\\ [1mm]
49 & - & H13-3 & - & 0:54:59.92 & -37:41:50.68 & 26.08$\pm$0.07 & - & - & - \\ [1mm]
50 & - & H14-14 & PN56 & 0:54:57.22 & -37:41:22.89 & 26.09$\pm$0.07 & 0.00\tablefootmark{*} & - & -\\ [1mm]
51 & A-29 & H14-6 & - & 0:54:55.72 & -37:41:26.39 & 26.12$\pm$0.06& 0.00\tablefootmark{*} & - & -\\ [1mm]
52 & - & H17-5 & - & 0:54:49.75 & -37:42:59.50 & 26.12$\pm$0.06 & - & - & -\\ [1mm]
53 & - & H2-1 & - & 0:54:53.10 & -37:39:34.10 & 26.17$\pm$0.07 & -  & - & - \\ [1mm]
54 & - & H16-5 & - & 0:54:52.90 & -37:44:00.13 & 26.18$\pm$0.08 & 0.00\tablefootmark{*} & - & - \\ [1mm]
55 & J-1 & L10-3 & - & 0:54:50.62 & -37:41:46.48 & 26.19$\pm$0.07& - & - & -\\ [1mm]
56 & - & L9-6 & PN47 & 0:54:54.36 & -37:42:32.71 & 26.24$\pm$0.07 & - & - & -\\ [1mm]
57 & - & H16-2 & PN50 & 0:54:55.00 & -37:43:11.53 & 26.28$\pm$0.08& - & - & -\\ [1mm]
58 & - & H10-5 & - & 0:54:44.42 & -37:38:49.86 & 26.35$\pm$0.07& 0.00\tablefootmark{*} & - & - \\ [1mm]
59 & - & H11-1 & PN15 & 0:54:39.03 & -37:38:43.34 & 26.36$\pm$0.07 & 0.00\tablefootmark{*} & - & - \\ [1mm]
60 & - & H14-9 & - & 0:54:54.27 & -37:41:35.21 & 26.38$\pm$0.07& -& - & - \\ [1mm]
61 & - & L12-4 & PN18 & 0:54:39.81 & -37:41:34.87 & 26.48$\pm$0.09& - & - & -\\ [1mm]
62 & D-7 & L12-5 & PN11 & 0:54:37.74 & -37:41:18.90 & 26.50$\pm$0.06& - & - & -\\ [1mm]
63 & - & H2-4 & - &  0:54:54.52 & -37:39:12.97 & 26.53$\pm$0.07 & - & - & -\\ [1mm]
64 & - & H2-3 &- & 0:54:54.58 & -37:39:06.09 & 26.56$\pm$0.07& - & - & -\\ [1mm]
65 & - & L6-11 & - & 0:54:43.66 & -37:40:10.98 & 26.60$\pm$0.10& - & - & -\\ [1mm]
66 & - & H10-6 & - & 0:54:45.69 & -37:38:58.76 & 26.61$\pm$0.09 & - & - & -\\ [1mm]
67 & - & H13-7 & - & 0:54:57.08 & -37:42:03.65 & 26.64$\pm$0.08& - & - & -\\ [1mm]
68 & A-24 & L4-17 & - & 0:54:52.16 & -37:41:32.42 & 26.73$\pm$0.07 & - & - & -\\ [1mm]
69 & E-5 & L6-17 & PN19 & 0:54:40.02 & -37:40:02.33 & 26.74$\pm$0.06 & - & - & -\\ [1mm]
70 & - & L2-9 & - & 0:55:04.57 & -37:40:30.39 & 26.77$\pm$0.07 & - & - & -\\ [1mm]
71 & - & H8-13 & - & 0:54:51.58 & -37:39:54.83 & 26.85$\pm$0.08& - & - & -\\ [1mm]
72 & - & L2-12 & - & 0:55:04.30 & -37:40:24.76 & 26.88$\pm$0.08& -  & - & -\\ [1mm]
73 & A-7 & H14-3 & - & 0:54:54.56 & -37:41:10.83 & 26.90$\pm$0.07& - & - & -\\ [1mm]
74 & - & H14-12 & - & 0:54:54.22 & -37:41:50.64 & 26.94$\pm$0.08& - & - & -\\ [1mm]
75 & - & H12-12 & - & 0:55:03.12 & -37:42:26.65 & 26.94$\pm$0.09& 0.00\tablefootmark{*} & - & - \\ [1mm]
76 & - & H1-5 & - & 0:54:58.93 & -37:39:46.00 & 27.04$\pm$0.08& - & - & -\\ [1mm]
77 & - & H9-7 & - & 0:54:47.73 & -37:39:31.07 & 27.06$\pm$0.08& - & - & -\\ [1mm]
78 & J-4 & L10-8 & - & 0:54:48.98 & -37:42:09.55 & 27.10$\pm$0.07& - & - & -\\ [1mm]
79 & - & H13-8 & - & 0:55:02.07 & -37:42:06.89 & 27.13$\pm$0.10& - & - & -\\ [1mm]
80 & - & H13-4 & - & 0:55:01.30 & -37:42:19.86 & 27.15$\pm$0.10& - & - & -\\ [1mm]
81 & - & H8-7 & - & 0:54:53.07 & -37:39:45.15 & 27.26$\pm$0.09& - & - & -\\ [1mm]
82 & - & H13-1 & - & 0:54:59.75 & -37:41:35.90 & 27.29$\pm$0.11& - & - & -\\ [1mm]
83 & C-11 & - & - & -0:54:45.86 & -37:40:46.33 & 27.34$\pm$0.07 & - & - & -\\ [1mm]
84 & - & H17-3 & - & 0:54:51.91 & -37:43:52.93 & 27.34$\pm$0.10 & - & - & -\\ [1mm]
85 & B-19 & L11-4 & - & 0:54:47.24 & -37:41:19.98 & 27.50$\pm$0.07 & - & - & -\\ [1mm]
86 & D-9 & - & - & 0:54:38.99 & -37:41:10.65 & 27.56$\pm$0.07& - & - & -\\ [1mm]
87 & - & H12-17 & - & 0:55:06.27 & -37:42:07.55 & 27.58$\pm$0.11 & - & - & -\\ [1mm]
88 & - & H17-8 & - & 0:54:50.96 & -37:42:50.68 & 27.59$\pm$0.11 & - & - & -\\ [1mm]
89 & - & H15-5 & - & 0:54:56.36 & -37:42:39.91 & 27.62$\pm$0.13 & - & - & -\\ [1mm]
90 & J-14 & - & - & 0:54:46.56 & -37:42:29.36 & 27.67$\pm$0.09 & - & - & -\\ [1mm]
91 & - & L1-7 & - & 0:55:07.22 & -37:40:44.58 & 27.67$\pm$0.12 & - & - & -\\ [1mm]
92 & E-1 & - & - & 0:54:41.25 & -37:39:40.11 & 27.69$\pm$0.08 & -& - & -\\ [1mm]
93 & E-9 & - & - & 0:54:38.85 & -37:39:58.80 & 27.73$\pm$0.08& - & - & -\\ [1mm]
94 & J-8 & - & - & 0:54:46.53 & -37:41:55.03 & 27.75$\pm$0.10& - & - & -\\ [1mm]
95 & D-19 & - & - & 0:54:38.89 & -37:40:38.47 & 27.76$\pm$0.08& - & - & -\\ [1mm]
96 & - & H11-3 & - & 0:54:40.57 & -37:38:28.75 & 27.80$\pm$0.13& - & - & -\\ [1mm]
97 & - & H2-5 & - & 0:54:57.74 & -37:39:12.84 & 27.86$\pm$0.13 & -& - & -\\ [1mm]
98 & C-14 & - & - & 0:54:42.26 & -37:41:21.15 & 27.88$\pm$0.10 & - & - & -\\ [1mm]
99 & - & L2-10 & - & 0:55:03.98 & -37:40:06.30 & 27.90$\pm$0.13 & -& - & -\\ [1mm]
100 & A-6 & - & - & 0:54:52.56 & -37:40:45.39 & 28.02$\pm$0.10 & - & - & -\\ [1mm]
101 & - & H12-15 & - & 0:55:04.64 & -37:41:59.55 & 28.06$\pm$0.16 & - & - & -\\ [1mm]
102 & I-7 & - & - & 0:54:42.12 & -37:42:14.90 & 28.32$\pm$0.12 & - & - & -\\ [1mm]
103 & A-15 & - & - & 0:54:53.80 & -37:41:30.34 & 28.40$\pm$0.13 & - & - & -\\ [1mm]
104 & A-52 & - & - & 0:54:51.93 & -37:41:26.37 & 28.62$\pm$0.15 & - & - & -\\ [1mm]
105 & E-7 & - & - & 0:54:40.27 & -37:39:56.77 & 28.62$\pm$0.12 & - & - & -\\ [1mm]
106 & E-13 & - & - & 0:54:40.56 & -37:40:01.23 & 28.84$\pm$0.13 & - & - & -\\ [1mm]
107 & B-39 & - & - & 0:54:46.25 & -37:40:49.75 & 28.91$\pm$0.15 & - & - & -\\ 

\end{longtable}
\tablefoot{\\
\tablefoottext{a}{Lower limits assuming maximum \oiii~conversion efficiency of 11\%}\\
\tablefoottext{*}{Lower limit value}
}
}

\end{appendix}
\end{document}